\documentclass[preprint,showpacs,preprintnumbers,amsmath,amssymb]{revtex4}

\usepackage{graphicx}% Include figure files
\usepackage{dcolumn}% Align table columns on decimal point
\usepackage{bm}% bold math

\begin{document}

\preprint{}

\title{Extended Studies of Separability Functions and Probabilities and 
the Relevance of Dyson Indices}

\author{Paul B. Slater}% 
\email{slater@kitp.ucsb.edu}
\affiliation{%
ISBER, University of California, Santa Barbara, CA 93106\\
}%
\date{\today}% It is always \today, today,

\begin{abstract}
We report substantial progress in the
study of {\it separability functions}
and their application to the computation of {\it separability 
probabilities} for the real, complex and quaternionic qubit-qubit 
and qubit-qutrit systems. We expand our recent work 
({\it J. Phys. A} {{\bf 39}}, 913 [2006]), in which the Dyson indices 
of random matrix theory played an essential role,  
to include the use of 
not only the volume element of the 
Hilbert-Schmidt (HS) 
metric, but also that of the Bures (minimal monotone)
metric as measures over these finite-dimensional quantum 
systems. Further, we now employ the 
Euler-angle parameterization of density matrices ($\rho$), in addition to
the Bloore parameterization.
The Euler-angle separability function for the minimally degenerate 
complex two-qubit states is well-fitted by the sixth-power of 
the participation ratio, $R(\rho)=\frac{1}{\mbox{Tr} \rho^2}$.
Additionally, replacing $R(\rho)$ by
a simple linear transformation of the 
Verstraete-Audenaert-De Moor function ({\it Phys. Rev. 
A}, {\bf{64}}, 012316 [2001]), 
we find close adherence
to Dyson-index behavior for the  
real and complex (nondegenerate) two-qubit scenarios.
Several of the analyses reported help to 
fortify our conjectures that
the HS and Bures separability probabilities  
of the complex two-qubit
states are $\frac{8}{33} \approx 0,242424$ and
$\frac{1680 (\sqrt{2}-1)}{\pi^8} \approx 0.733389$, respectively. 
Employing certain {\it regularized beta functions} in the role
of Euler-angle separability functions, we closely reproduce--consistently with the Dyson-index {\it ansatz}--several
HS two-qubit separability probability conjectures.

\end{abstract}


\maketitle
\tableofcontents
\section{Introduction}
For several years now, elaborating upon an idea proposed in 
\cite{ZHSL}, we have been pursuing the problem of deriving
(hypothetically exact) formulas for the proportion of states of 
qubit-qubit \cite{avron} and qubit-qutrit \cite{sudarshan} 
systems that are {\it separable} 
(classically-correlated) in nature 
\cite{slaterHall,slaterA,slaterC,slaterOptics,slaterJGP,slaterPRA,pbsCanosa,slaterPRA2}.
Of course, any such proportions will critically depend upon the measure 
that is placed upon 
the quantum 
systems. In particular, we have---in analogy to (classical) Bayesian
analyses, in which the {\it volume element} of the {\it Fisher information} 
metric for a parameterized family of probability distributions 
is utilized as a measure (``Jeffreys' prior'') 
\cite{kass}---principally 
employed the volume elements of the well-studied (Euclidean, flat) 
Hilbert-Schmidt (HS) and 
Bures ({\it minimal} monotone or symmetric-logarithmic-derivative [SLD]) 
metrics (as well as a number of other 
[non-minimal] {\it monotone} metrics \cite{slaterJGP}).

\.Zyczkowski and Sommers \cite{szHS,szBures} 
have, using methods of random matrix theory \cite{random}
(in particular, the Laguerre ensemble), obtained formulas,
general for all $n$, for the 
HS and Bures {\it total} volumes (and hyperareas) of $n \times n$
(real and complex)
quantum systems. Up to normalization factors, the HS total volume 
formulas were also found by Andai \cite{andai}, in a rather different 
analytical framework, using a number of 
(spherical and beta) integral identities 
and positivity (Sylvester) conditions. (He also obtained 
formulas---general for any monotone metric [including the Bures]---for the 
volume of {\it one}-qubit [$n=2$] states \cite[sec. 4]{andai}.)

Additionally, Andai did specifically study 
the HS {\it quaternionic} case. He derived the
HS total volume for $n \times n$ quaternionic systems \cite[p. 13646]{andai},
\begin{equation} \label{andaiQuatVol0} 
V_{quat}^{HS} = \frac{(2 n-2)! 
\pi^{n^2-n}}{(2 n^2-n-1)!} \Pi_{i=1}^{n-2} (2 i)!,
\end{equation}
giving us for the two-qubit ($n=4$) case that will be our specific 
initial interest here,
the 27-dimensional volume,
\begin{equation} \label{andaiQuatVol}
\frac{\pi ^{12}}{7776000} \cdot \frac{1}{40518448303132800} =
\frac{\pi ^{12}}{315071454005160652800000}
\approx 2.93352 \cdot 10^{-18}.
\end{equation}
(In the analytical setting employed by \.Zyczkowski and Sommers \cite{szHS}, 
this volume would appear as 
$2^{12}$ times as large \cite[p. 13647]{andai}.)

If one then possessed a
companion volume formula for the {\it separable} 
subset, one could immediately compute 
the HS two-qubit quaternionic separability {\it probability} 
($P^{HS}_{quat}$) by taking
the ratio of the two volumes. 
In fact, following a convenient paradigm we have developed, and will employ
several times below, in varying contexts, we will 
compute $P^{HS}_{quat}$ as the product ($R_1 R_2$) of two {\it ratios}, 
$R_1$ and $R_2$. The first (24-dimensional) factor on the left-hand side of 
(\ref{andaiQuatVol}) 
will serve
as the denominator of $R_1$ and the second (3-dimensional) 
factor, as the denominator of $R_2$. The determinations 
of the {\it numerators} 
of such pairs of complementary ratios will 
constitute, in essence, our (initial) principal computational
challenges.

\subsection{Bloore parameterization of density matrices} \label{secBloore}
One analytical approach to the separable volume/probability 
question that has 
recently proved to be productive \cite{slater833}---particularly, in 
the case of the Hilbert-Schmidt (HS) metric (cf. \cite{slaterDyson})---makes 
fundamental use of a (quite elementary) 
form of density matrix parameterization first
proposed by Bloore \cite{bloore}. This methodology 
can be seen to be strongly related
to the very common and long-standing use of {\it correlation matrices} 
in statistics and its many fields of application \cite{joe,kurowicka,kurowicka2}.
(Correlation matrices can be obtained by standardizing {\it covariance} 
matrices. Density matrices have been viewed as covariance matrices of
multivariate normal [Gaussian] distributions \cite{guiasu}. Covariance 
matrices for certain observables 
have been used to study the separability of finite-dimensional 
quantum systems \cite{guhne}. The possible 
states of polarization of a two-photon system are describable by six
Stokes parameters and a $3 \times 3$ ``polarization correlation'' matrix 
\cite{vanik}.)

In the Bloore (off-diagonal scaling)
parameterization, one simply represents an off-diagonal {\it ij}-entry
of a density matrix $\rho$, as $\rho_{ij} = \sqrt{\rho_{ii} \rho_{jj}} w_{ij}$,
where $w_{ij}$ might be real, complex or quaternionic 
\cite{asher2,adler,batle2} in nature.
The particular attraction of the Bloore scheme, in terms of the 
separability problem in which we are interested, is that one can 
(in the two-qubit case) implement
the well-known Peres-Horodecki separability (positive-partial-transpose) 
test \cite{asher,michal} 
using only the ratio,
\begin{equation} \label{firstratio}
\mu =\sqrt{\nu} = \sqrt{\frac{\rho_{11} \rho_{44}}{\rho_{22} 
\rho_{33}}},
\end{equation}
 rather than the four (three independent) 
diagonal entries of $\rho$ individually \cite[eq. (7)]{slaterPRA2} 
\cite[eq. (5)]{slater833}.

Utilizing the Bloore parameterization, we have, accordingly, been able to reduce the problem
of computing the desired HS volumes of two-qubit separable states 
 to the computations 
of {\it one}-dimensional integrals (\ref{Vsmall}) 
over $\mu \in [0,\infty]$. The 
associated integrands are the 
{\it products} of
{\it two} 
functions, one a readily determined jacobian function 
$\mathcal{J}(\mu)$ (corresponding, first, 
to 
the transformation to the Bloore variables $w_{ij}$ 
and, then, to $\mu$) 
 and the other, 
the more problematical (what we have termed) 
{\it separability function} $\mathcal{S}^{HS}(\mu)$ 
\cite[eqs. (8), (9)]{slaterPRA2}.

In the qubit-{\it qutrit} case [sec.~\ref{QubQut}], {\it two} 
ratios,
\begin{equation} \label{tworatios}
\nu_{1}= \frac{\rho_{11} \rho_{55}}{\rho_{22} \rho_{44}}, \hspace{.2in}
\nu_{2}= \frac{\rho_{22} \rho_{66}}{\rho_{33} \rho_{55}},
\end{equation}
are required to express the separability conditions (choosing to compute
the partial transpose by transposing four $3 \times 3$ blocks, rather than
nine $2 \times 2$ blocks of the $6 \times 6$ density matrices),, 
but analytically the corresponding HS separability
functions also appear to be {\it univariate} in nature, being 
simply functions of either $\nu_1$ or of $\nu_2$ singly, or 
the product \cite[sec. III]{slater833},
\begin{equation} 
\eta = \nu_1 \nu_2 =\frac{\rho_{11} \rho_{66}}{\rho_{33} \rho_{44}}.
\end{equation}
\subsection{Euler-angle parameterization of density matrices}
Here, one can again divide the set of parameters into two groups, in a
natural manner (that is, 
the diagonal and off-diagonal parameters in the Bloore
framework). Now, the two sets are composed of the eigenvalues of $\rho$ and
of the Euler angles parameterizing the associated unitary matrix of
eigenvectors \cite{sudarshan,tbs}.

With both forms of parameterizations we have discussed, 
one can obtain the total volume of $n \times n$ quantum
systems as the {\it product} of integrals over the two complementary sets 
\cite{szHS,szBures}.
But this direct approach 
no longer holds in terms of computing the separable volume.
So, we have evolved the following general strategy 
\cite{slaterPRA2,slater833}. 
We integrate over the larger set (off-diagonal or Euler-angle
parameters), {\it while} enforcing separability conditions, leaving us
with {\it separability functions} that are functions of only the {\it smaller}
set of parameters (diagonal entries or eigenvalues). Doing so, of course,
substantially reduces the dimensionality of the problem.

We are, then, left with such separability functions and the 
ensuing task of 
appropriately integrating these functions
 over the remaining parameters (diagonal
entries or eigenvalues), so as to obtain the requisite {\it separable
volumes}.
\subsection{Immediately preceding studies}
In our extensive numerical (quasi-Monte Carlo integration) 
investigation \cite{slaterPRA2} of the 9-dimensional and 15-dimensional 
convex sets of real and complex $4 \times 4$ density matrices,
we had formulated ans{\"a}tze for the two associated separability 
functions ($\mathcal{S}^{HS}_{real}(\mu)$ and $\mathcal{S}^{HS}_{complex}(\mu)$), 
proposing that
they were 
proportional to certain (independent) 
{\it incomplete beta functions} \cite{handbook},
\begin{equation}
B_{\mu^2}(a,b) =\int_{0}^{\mu^2} \omega^{a-1} (1-\omega)^{b-1} d \omega,
\end{equation}
for particular values of $a$ and $b$.
However, in 
the subsequent study 
\cite{slater833}, 
we were led to somewhat 
modify these ans{\"a}tze, in light of multitudinous 
exact {\it lower}-dimensional results obtained there. Since
these further results clearly manifested patterns fully consistent with 
the {\it Dyson index} (``repulsion exponent'') 
pattern ($\beta =1, 2, 4$) 
of random matrix theory \cite{dyson}, we proposed
that, in the (full 9-dimensional) real case,
the separability function was proportional to a specific 
incomplete 
beta 
function ($a=\frac{1}{2},b=2$),
\begin{equation}
\mathcal{S}^{HS}_{real}(\mu) \propto B_{\mu^2}(\frac{1}{2},2) 
\equiv  \frac{3}{4} (3 -\mu^2) \mu 
= \frac{3|\rho|^{\frac{1}{2}}}
{4 \rho_{22} \rho_{33}} (3 \rho_{22} \rho_{33}- \rho_{11} \rho_{44}) 
\end{equation}
and in the complex case, proportional, not just 
to an independent function, but 
simply to the {\it square} of
$\mathcal{S}^{HS}_{real}(\mu)$.
(These proposals are strongly consistent 
\cite[Fig. 4]{slater833} with the numerical
results generated in \cite{slaterPRA2}.) This 
chain of reasoning, then, 
immediately suggests the further proposition 
that the separability function in
the {\it quaternionic} case is exactly proportional to the {\it fourth}
power of that for the real case (and, obviously, the square of that for
the complex case). It is that specific proposition we will, first, 
seek to evaluate here.
\subsection{Objectives}
We seek below (sec.~\ref{secQuat}) to further test the validity of our 
Dyson-index ansatz, first advanced in \cite{slater833}, 
as well as possibly develop an enlarged perspective
on the still not yet fully 
resolved problem of the two-qubit 
HS separability probabilities 
in all three (real, complex and quaternionic) cases.
(In \cite{slater833}, we proposed, combining numerical and theoretical 
arguments--not fully rising to the level of a formal
demonstration--that in the real two-qubit case, the HS separability probability
is $\frac{8}{17}$, and in the complex 
two-qubit case,  $\frac{8}{33}$.) A supplementary treatment of
the {\it truncated} quaternionic scenario ($\beta=3$) is presented in 
sec.~\ref{secTrunc}. 

We invesigate related separability-function questions 
 in the qubit-{\it qutrit} framework, again making use of
the Hilbert-Schmidt metric (sec.~\ref{QubQut}), and, 
also in the two-qubit setting, employing the Bures (minimal monotone) metric 
(sec.~\ref{secBures}). 
Since it becomes more problematical to obtain separability functions 
in the Bures case, we explore--as originally proposed in \cite{slaterJPAreject}--the use of the $SU(4)$ Euler-angle parameterization of Tilma, Byrd and 
Sudarshan \cite{tbs} for similarly-minded purposes (sec.~\ref{secEuler}).
\section{Bloore-parameterization separability functions} \label{Bp}
\subsection{Quaternionic two-qubit Hilbert-Schmidt analysis} 
\label{secQuat}
Due to the ``curse of dimensionality'' \cite{bellman,kuosloan}, 
we must anticipate that for the
same number of sample ("low-discrepancy" Tezuka-Faure (TF) 
\cite{giray1,tezuka})
points generated in the quasi-Monte
Carlo integration 
procedure employed in \cite{slaterPRA2} and here, our numerical estimates
of the quaternionic separability function will be less precise than 
the estimates  were for
the complex, and {\it a fortiori}, real cases. 
(An interesting, sophisticated alternative
approach to computing the Euclidean volume 
of {\it convex} bodies involves a variant of 
{\it simulated annealing} \cite{lovasz} (cf. \cite{dyer}), and allows one---unlike the Tezuka-Faure approach, we have so far employed---to 
establish confidence intervals for estimates.)

Our first extensive numerical 
analysis here involved the generation of sixty-four  million 24-dimensional
Tezuka-Faure points, all situated in 
the 24-dimensional unit hypercube $[0,1]^{24}$.
(The three independent 
diagonal entries of the density matrix $\rho$---being incorporated 
into the jacobian $\mathcal{J}(\mu)$---are
irrelevant at this stage of the calculations 
of $\mathcal{S}^{HS}_{quat}(\mu)$. 
The 24 [off-diagonal] Bloore variables had been linearly
transformed so that each ranged over the unit interval [0,1]. 
The computations were done over several weeks, using compiled Mathematica 
code, on a MacMini workstation.) 

Of the sixty-four million sample points 
generated, 7,583,161, approximately 12$\%$, 
 corresponded to possible
$4 \times 4$ quaternionic density matrices---satisfying nonnegativity 
requirements. For each of these feasible points, we evaluated whether or not
the Peres-Horodecki positive-partial-transpose separability test was 
satisfied for 2,001 equally-spaced values of $\mu \in [0,1]$.

Here, we encounter another computational ``curse'', 
in addition to that already 
mentioned pertaining to the high-dimensionality of our problem, 
and also the infeasibility 
of most ($88\%$) of the sampled Tezuka-Faure points. In the standard 
manner \cite[eq. (5.1.4)]{random} 
\cite[p. 495]{adler} \cite[eq. (17)]{slaterJMP1996}  
\cite[sec. II]{JIANG}, 
making use of 
the Pauli matrices, we 
transform the $4 \times 4$ {\it quaternionic} density matrices---and their 
partial transposes---into 
$8 \times 8$ density matrices with [only] complex entries. 
Therefore, given a feasible 24-dimensional point, we have to check
for each of the 2,001 values of $\mu$, an $8 \times 8$
matrix for nonnegativity, rather than a $4 \times 4$ one, as was done in
both the real and complex two-qubit cases. In all three of these cases, 
we found that it would be incorrect
to simply assume---which would, of course, speed computations---that 
if the separability test is passed for a certain $\mu_{0}$, 
it will also be passed for all $\mu$ lying between $\mu_{0}$ and 1. 
This phenomenon reflects the intricate (quartic 
{\it both} in $\mu$ and in the Bloore variables $w_{ij}$'s, 
in the real and complex cases) nature of the
polynomial separability constraints 
\cite[eq. (7)]{slaterPRA2} \cite[eq. (5)]{slater833}.
\subsubsection{Estimated separability function and probability}
\begin{figure}
\includegraphics{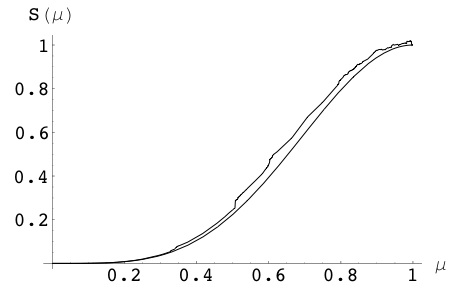}
\caption{\label{fig:quatsepfunct}Estimate---based on 64,000,000 sampled 
24-dimensional points---of the normalized form 
of the two-qubit {\it quaternionic} 
separability function $S^{HS}_{quat}(\mu)$, along with its (well-fitting) 
hypothetical true form, 
the {\it fourth} power of the normalized
form of $\mathcal{S}^{HS}_{real}(\mu)$, that is,
$\Big(\frac{1}{2} (3 -\mu^2) \mu\Big)^4$}
\end{figure}

In Fig.~\ref{fig:quatsepfunct} we show the estimate we, thus, were able 
to obtain
of the two-qubit quaternionic separability function 
$\mathcal{S}^{HS}_{quat}(\mu)$, in its normalized form. 
(Around $\mu=1$, one must have the evident symmetrical relation 
$\mathcal{S}^{HS}(\mu) = \mathcal{S}^{HS}(\frac{1}{\mu})$.) Accompanying our estimate 
in the plot is
the  (well-fitting) hypothetical true
form (according with our Dyson-index ansatz 
\cite{slater833}) of the HS two-qubit separability function, 
that is, the {\it fourth} power, $\Big(\frac{1}{2} (3 -\mu^2) \mu\Big)^4$, of the 
normalized form of $\mathcal{S}^{HS}_{real}(\mu)$.

For the specific, important value of 
$\mu=1$--implying that $\rho_{11} 
\rho_{44}=\rho_{22} \rho_{33}$--the ratio ($R_{1}$) of the 24-dimensional 
HS measure ($m_{sep} = R^{numer}_{1}$) 
assigned in our 
estimation procedure to separable
density matrices to the (known) total 24-dimensional 
HS measure ($m_{tot} =R^{denom}_{1}$) 
allotted to all (separable and 
nonseparable) density matrices is $R_{1} = 0.123328$. 
The exact value of $m_{sep}$ is, of course, to begin here, unknown, being
a principal desideratum of our investigation. On the other hand, 
we can directly deduce that 
$m_{tot} = R_{1}^{denom} = 
\frac{\pi ^{12}}{7776000} \approx 0.118862$---our sample 
estimate  being 0.115845---by dividing the two-qubit HS quaternionic 
27-dimensional volume 
(\ref{andaiQuatVol}) obtained by Andai \cite{andai} by
\begin{equation} \label{rationalfraction}
R_{2}^{denom} =2 \int_{0}^{1} \mathcal{J}_{quat}(\mu) d \mu =
\frac{\Gamma \left(\frac{3 \beta }{2}+1\right)^4}{\Gamma
   (6 \beta +4)} =
\frac{1}{40518448303132800} \approx 2.46801 \cdot 10^{-17}, 
\hspace{.1in} \beta  = 4.
\end{equation}

\begin{figure}[!tbp]
\includegraphics{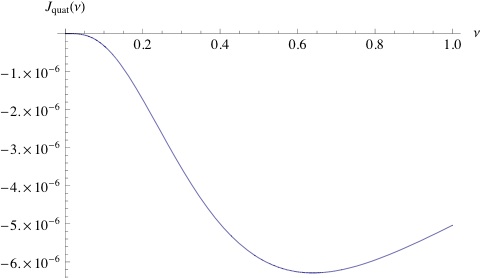}
\caption{\label{fig:quatjacobian}The univariate quaternionic jacobian function 
$\mathcal{J}_{quat}(\mu)$}
\end{figure}

Here, $\mathcal{J}_{quat}(\mu)$ is the quaternionic jacobian function
 (Fig.~\ref{fig:quatjacobian}),
obtained by transforming the
quaternionic Bloore
jacobian $\Big(\rho_{11} \rho_{22} \rho_{33} 
(1-\rho_{11} -\rho_{22} -\rho_{33})\Big)^\frac{3 \beta}{2}$, $\beta=4$, 
to the $\mu$ variable by replacing, say $\rho_{33}$ by $\mu$,
and integrating out $\rho_{11}$ and $\rho_{22}$.
(We had presented plots of $\mathcal{J}_{real}(\mu)$ 
and $\mathcal{J}_{complex}(\mu)$ 
in \cite[Figs. 1, 2]{slaterPRA2}, and observed 
apparently highly oscillatory behavior in both functions in 
the vicinity of $\mu=1$. 
However, a referee of \cite{slater833} informed us that this was simply
an artifact of using standard machine precision, and that with 
sufficiently enhanced
precision [only recently available for plotting purposes in Mathematica 
6.0]--as now employed in Fig.~\ref{fig:quatjacobian}--the 
oscillations could be seen to be, in fact, illusory.)
We theoretically can obtain the two-qubit 
quaternionic {\it separability} 
probability $P^{HS}_{sep/quat}$ by multiplying the true value 
(which we do not beforehand know, but seek) of the ratio
$R_{1}$ by a second (known, computable) ratio $R_{2}$.
The denominator of $R_{2}$ has already been given
(\ref{rationalfraction}).
The {\it numerator} of $R_{2}$ is the specific value
\begin{equation} \label{R2numerator}
R_{2}^{numer} = 2 \int_{0}^{1} \mathcal{J}_{quat}(\mu) 
\Big(\frac{1}{2} (3 -\mu^2) \mu\Big)^4 d \mu = \frac{5989}{358347086242825680000} 
\approx 1.67128 \cdot 10^{-17},
\end{equation}
where, to obtain the integrand, 
we have multiplied (in line with our basic 
[Bloore-parameterization] approach to the separability 
probability question) 
the quaternionic jacobian function (Fig.~\ref{fig:quatjacobian}) 
by the 
(normalized) putative 
form of the two-qubit quaternionic separability function. 
(Note the use of the $\beta=4$ exponent.)

The counterpart of $R^{numer}_{2}$ in the 9-dimensional {\it real} case is  
$\frac{1}{151200}$ and in the 15-dimensional {\it complex} case, 
$\frac{71}{99891792000}$. We further note, regarding the last 
denominator, that 
\begin{equation}
99891792000 = \left(
\begin{array}{c}
 \text{11} \\
 \text{2}
\end{array}
\right) \frac{\Gamma{(16)}}{\Gamma{(7)}}
\end{equation}
is the coefficient 
of $\mu^2$ in $11 !  
L_{11}^{4}(\mu)$ and $\frac{151200}{2}= 75600$ plays the exact same role 
in $6! L_{6}^{4}(\mu)$, where $L_{m}^{4}(\mu)$ is a generalized ($a=4$)
Laguerre polynomial  (see sequences A062260 and A062140  
in the {\it The On-Line Encylopaedia of Integer
Sequences}). (Also, as regards the denominator of (\ref{R2numerator}), 
$\frac{358347086242825680000}{3587352665} = 99891792000$.)
\.Zyczkowski and Sommers had 
made use of the Laguerre ensemble in deriving
the HS and Bures volumes and hyperareas of $n$-level quantum systems 
\cite{szHS,szBures}. Generalized (associated/Sonine) Laguerre 
polynomials [``Laguerre functions''] have been employed
in another important quantum-information context, in 
proofs of Page's conjecture on the average entropy of a subsystem 
\cite{sanchez,sen}.)

We, thus, have, for our two-qubit quaternionic case, that
\begin{equation} \label{ratio2}
R_{2} = \frac{R_{2}^{numer}}{R_{2}^{denom}} = 
\frac{125769}{185725} \approx 0.677179.
\end{equation}
(The {\it real} counterpart of $R_{2}$ is
$\frac{1024}{135 \pi ^2} \approx 0.76854$, and the 
{\it complex} one, $\frac{71}{99} \approx 0.717172$. 
Additionally, we computed that the 
corresponding ``truncated'' quaternionic \cite{pfaff}
ratio---when {\it one} of the four quaternionic parameters is set to 
zero, that is the Dyson-index 
case $\beta=3$--- is $\frac{726923214848}{106376244975 
\pi ^2} \approx 0.692379$. Thus, we see that these four 
important ratios $R_{2}(\beta)$
monotonically 
decrease as $\beta$ increases, and also, significantly, that the two 
ratios for odd values of $\beta$ 
differ qualitatively---both having $\pi^2$ in their denominators---from 
those two for even $\beta$.)

Our quasi-Monte Carlo (preliminary)
estimate of the two-qubit quaternionic separability 
{\it probability} is, then, 
\begin{equation} \label{r1r2}
P_{sep/quat}^{HS} \approx R_{1} R_{2} =0.0813594.
\end{equation}
Multiplying the total volume of the 27-dimensional convex set of 
two-qubit quaternionic states, given in the framework 
of Andai \cite{andai} by (\ref{andaiQuatVol}), by this result 
(\ref{r1r2}), we 
obtain the two-qubit quaternionic separable volume 
estimate $V^{HS}_{sep/quat} \approx  2.38775 \cdot 10^{-19}$. 

Our 24-dimensional quasi-Monte Carlo integration procedure 
leads to a derived estimate of (the total 27-dimensional volume) 
$V^{HS}_{quat}$, that was
somewhat smaller, $2.85906
\cdot 10^{-18}$, than the actual value $2.93352 \cdot 10^{-18}$ 
given by (\ref{andaiQuatVol}). 
Although rather satisfying, this 
was sufficiently imprecise to discourage us
from further attempting to ``guestimate'' the 
(all-important) constant ($R_{1}$) by which to multiply
the putative normalized form, $(\frac{1}{2} (3-\mu^2) \mu)^4$, 
of the quaternionic separability function in (\ref{R2numerator}) 
in order to yield
the true separable volume.
In our previous study \cite[sec. IX.A]{slater833}, we presented certain plausibility
arguments to the effect that
the corresponding $R_{1}$
constant in the 9-dimensional real case was
$\frac{135 \pi^2}{2176} = (\frac{20 \pi^4}{17})/(\frac{512 \pi^2}{27})$, and 
$\frac{24}{71} =(\frac{256 \pi^6}{639})/(\frac{32 \pi^6}{27})$ in the 15-dimensional complex case.
(This leads---multiplying by the corresponding $R_{2}$'s, 
$\frac{1024}{135 \pi^2}$ and $\frac{71}{99}$---to 
separability probabilities of $\frac{8}{17}$ and 
$\frac{8}{33}$, respectively.)
\subsubsection{Supplementary estimation of $R_{1}$ constant} \label{supp1}
In light of such imprecision, in our initial estimates, 
we undertook a supplementary 
analysis, in which, instead of examining each feasible 24-dimensional 
TF point
for 2,001 possible values of $\mu$, with respect to separability or not, 
we simply used $\mu=1$. This, of course,
allows us to significantly increase the number of 
points generated from the 64,000,000 so far employed. 

We, thusly, generated 1,360,000,000 points, finding 
that we obtained a remarkably good fit to the important ratio
$R_{1}$ of the 24-dimensional measure ($m_{sep}$), 
at $\mu=1$, assigned to the separable
two-qubit quaternionic density matrices to the (known) 
measure 
($m_{tot}=\frac{\pi^{12}}{7776000}$) by setting $R_{1} =(\frac{24}{71})^2 
\approx 0.114263$ (our sample estimate of this quantity 
being the very close 0.114262).  This is 
{\it exactly} the square of the corresponding ratio $\frac{24}{71}$ we had
conjectured (based on extensive numerical and theoretical evidence) for 
the full (15-dimensional) complex two-qubit case
in \cite{slater833}.
\subsubsection{Conjectured complex and quaternionic 
separability functions and probabilities}
Under this  hypothesis on $R_{1}$ for $\beta=4$,  we have the ensuing 
string of relationships
\begin{equation} \label{bigfish}
\mathcal{S}^{HS}_{quat}(\mu) = \Big(\frac{24}{71})^2 (\frac{1}{2} (3 -\mu^2) \mu\Big)^4 
=  \Big(\frac{6}{71}\Big)^2 \Big( (3 -\mu^2) \mu\Big)^4
=\Big( \mathcal{S}^{HS}_{complex}(\mu) \Big)^2,
\end{equation}
with (as already advanced in \cite{slater833}),
\begin{equation}
\mathcal{S}^{HS}_{complex}(\mu) = 
\frac{24}{71} \Big(\frac{1}{2} 
(3-\mu^2) \mu\Big)^2= \frac{6}{71} \Big((3-\mu^2) \mu\Big)^2.
\end{equation}
Then, using our knowledge of the complementary ratio $R_{2}$, given in 
(\ref{ratio2}), we obtain the desired exact result,
\begin{equation} \label{HSquat}
P^{HS}_{sep/quat} = R_{1} R_{2} = \frac{72442944}{936239725} 
=\frac{2^6 \cdot 3^3 \cdot 7 \cdot 53 \cdot 113}{5^2 \cdot 17 \cdot 19 \cdot 23 \cdot 71^2}
\approx 0.0773765,
\end{equation}
(the complex counterpart being $\frac{8}{33}$), 
as well as---in the framework of Andai \cite{andai}---that
\begin{equation}
 V^{HS}_{sep/quat} =\frac{5989 \pi ^{12}}{24386773433626137413880000000} 
\approx 2.26986 \cdot 10^{-19}.
\end{equation}
\subsection{{\it Truncated} quaternionic analysis ($\beta=3$)} \label{secTrunc}
For possible further insight into the HS two-qubit separability 
probability question,
we undertook a parallel quasi-Monte Carlo (Tezuka-Faure) integration 
(setting $\mu=1$) for
the truncated quaternionic case ($\beta=3$), in which one of the four
quaternionic parameters is set to zero. Although there was 
no corresponding formula
for the HS total volume for this scenario given in \cite{andai}, 
upon request, A. Andai 
kindly derived the result
\begin{equation} \label{puzzling}
V^{HS}_{trunc} = \frac{\pi ^{10}}{384458588946432000} 
\approx 2.43584 \cdot 10^{-13}.
\end{equation}
In fact, Andai was able to derive {\it one} simple 
overall comprehensive formula,
\begin{equation}
V^{HS}_{n,\beta}= \frac{\pi^{\frac{\beta n (n-1)}{4}}}{\Gamma (\beta \frac{n (n-1)}{2} +n)}  
\Pi_{i=1}^{n-1} \Gamma(\frac{i \beta}{2}+1)
\end{equation}
yielding
the total HS volumes for all $n \times n$ systems and Dyson indices $\beta$.
Let us, further, note that Andai obtains 
the result (\ref{puzzling}) as the product of three factors, 
$V^{HS}_{trunc} =\pi_1 \pi_2 \pi_3$, where
\begin{equation} \label{3factors}
\pi_1 = \frac{128 \pi ^8}{105}; \hspace{.1in} \pi_2 =\frac{128}{893025}; 
\hspace{.1in} 
\pi_3=\frac{189 \pi ^2}{12696335643836416}.
\end{equation}

Now, we will simply {\it assume}---in line with our basic Dyson-index 
ansatz, substantially supported in 
\cite{slater833} and above---that 
the corresponding separability function is of the form
\begin{equation}
\mathcal{S}^{HS}_{trunc}(\mu) \propto 
( (3 -\mu^2) \mu)^\beta, \hspace{.15in} \beta=3.
\end{equation}
(Of course, one should ideally {\it test} this 
specific application of the ansatz too, perhaps using the 
same quasi-Monte Carlo method we have 
applied to the $\beta =4$ instance above [Fig.~\ref{fig:quatsepfunct}].)

We were somewhat perplexed, however, by the results of our quasi-Monte Carlo
integration procedure, conducted in the 18-dimensional space of 
off-diagonal entries of the truncated quaterionic density matrix 
$\rho$. Though, we 
anticipated (from our previous 
extensive numerical experience here and elsewhere) that 
the estimate of the associated 18-dimensional volume would be, at least, 
within a few percentage points  of $\pi_1 \pi_2 = 
\frac{16384 \pi ^8}{93767625} \approx 
1.65793$, our actual 
estimate was, in fact, 
close to 0.967 (1, thus, falling within the possible margin of error).
Assuming the correctness of the analysis of Andai, which we have no other 
reason to doubt, the only possible explanations seemed to be that we had
committed some programming error (which 
we were unable to discern) or that we had
some conceptual misunderstanding regarding the analysis of truncated
quaternions. (Let us note that we do convert the $4 \times 4$ density matrix
to $8 \times 8$ [complex] form 
\cite[p. 495]{adler} \cite[eq. (17)]{slaterJMP1996}
\cite[sec. II]{JIANG}, while it appears that Andai does not 
directly employ such a transformation in his derivations.)

In any case, we did 
devote considerable computing time to the $\beta=3$ problem 
(generating 1,180,000,000 18-dimensional Tezuka-Faure 
points), with the hope being
that if we were in some way in error, the error would be 
an {\it unbiased} 
one, and 
that the all-important {\it ratio} of separable to total volume 
would be unaffected.

Proceeding thusly, our best estimate 
({\it not} making use of the Andai result (\ref{puzzling}) for the present) 
of the HS separability probability 
was 0.193006. One interesting possible candidate exact value
is, then, $\frac{128}{633} = \frac{2^7}{3 \cdot 13 \cdot 17} 
\approx 0.193062$. (Note the presence of 128 in the numerators, also, 
of both factors 
$\pi_1$ and $\pi_2$, given in (\ref{3factors}).)
This would give us a 
counterpart [$\beta=3$] value for the ratio $R_{2}$ of 
$\frac{160446825 \pi ^2}{5679087616} \approx 0.278838$. 
In \cite{slater833}, we had asserted that, in the other 
odd $\beta=1$ case, the counterpart of $R_{2}$ was
$\frac{135 \pi ^2}{2176} \approx 0.612315$. (Multiplying this by 
$\frac{1024}{135 \pi^2}$ gave us the 
conjectured HS {\it real} two-qubit separability probability 
of $\frac{8}{17}$.)

So, let us say that although we believe we have successfully
resolved---though still not having formal proofs---the 
two-qubit Hilbert-Schmidt 
separability probability question for the $\beta =2$ and 4 
(complex and quaternionic) cases, the odd ($\beta =1, 3$) cases, in 
particular $\beta =3$, appear at this point to be 
more problematical.
\subsection{Real and complex Qubit-{\it Qutrit} Hilbert-Schmidt Analyses} 
\label{QubQut}
For qubit-qutrit systems, we have previously reported 
\cite[eq. (44)]{slater833}, 
following the lines of our (Bloore-parameterization-based) 
two-qubit analyses,
that rather than the use of one ratio variable $\mu$, in implementing the 
Peres-Horodecki positive-partial-transpose test for 
separability,  it is necessary
to employ two (corresponding specifically here to the case where 
the partial transpose is implemented by transposing the four $3 \times 3$ 
blocks of the $6 \times 6$ density matrix $\rho$ in place) variables, 
already  
presented in 
(\ref{tworatios}).

Once again, employing the Tezuka-Faure quasi-Monte Carlo methodology, we 
generated 133,545 30-dimensional  and 1,950,000 20-dimensional {\it feasible}
points, corresponding now 
to the off-diagonal Bloore parameters of $6 \times 6$
complex and real
density matrices, respectively. 
(Each analysis 
was run on a MacMini workstation
for a number of weeks.)
The much larger number of feasible Tezuka-Faure points generated in the 
real case was primarily due to our reparameterization in that case 
of the Bloore off-diagonal entries (essentially correlations) 
in terms of {\it partial} correlations 
\cite{joe,kurowicka,kurowicka2} (cf. \cite{budden}). 
This allowed us to somewhat mitigate the computational ``curse'' 
of high dimensionality, in that {\it each} sampled point now corresponds to a 
density matrix and {\it none} (theoretically, at least) 
has to be discarded. (H. Joe has demonstated that it is 
possible to also implement this approach in the complex case, 
but the programming 
challenges for us were substantially 
greater, so we have not yet pursued such 
a course.) Of the 2,250,000 20-dimensional 
points, 38,622 were discarded because certain numerical difficulties 
(mainly convergence problems), arose in transforming to
the partial correlations. The 133,545 feasible 30-dimensional points 
were drawn from (a much larger) 430,000,000 ones.

\begin{figure}
\includegraphics{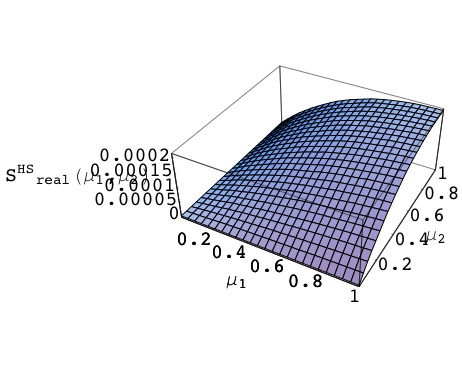}
\caption{\label{fig:QubQut}Interpolated estimate
over the unit square of the real qubit-qutrit
separability function 
$S_{real/qub-qut}^{HS}(\nu_{1},\nu_{2})$, based on 2,211,378 
20-dimensional Tezuka-Faure points. For {\it each} of these points, 
10,201 associated
$6 \times 6$ density matrices, parameterized by $\nu_1 \in [0,1]$ 
and $\nu_2 \in [0,1]$, were
tested for separability.}
\end{figure}

\begin{figure}
\includegraphics{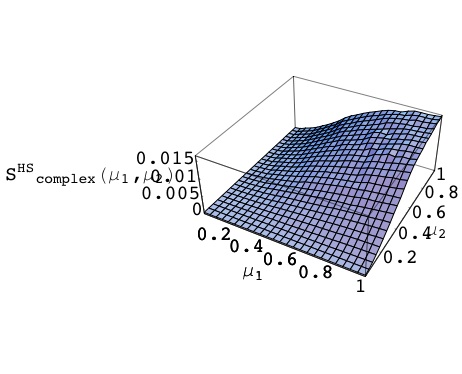}
\caption{\label{fig:QubQut2}Interpolated estimate
over the unit square of the complex qubit-qutrit
separability function $S_{complex/qub-qut}^{HS}(\nu_{1},\nu_{2})$, based on 133,545 
{\it feasible}
30-dimensional Tezuka-Faure points. For each point, 10,201 associated
$6 \times 6$ density matrices, parameterized by $\nu_1 \in [0,1]$ and 
$\nu_2\in [0,1]$, were
tested for separability.}
\end{figure}

For each feasible sampled point we tested whether the 
associated 
(real or complex) 
$6 \times 6$ density matrix was separable or not (that is, 
whether or not it passed the Peres-Horodecki test) for all possible 
pairs of $\nu_{1}$ and $\nu_{2}$ ranging from 0 to 1 in 
increments of $\frac{1}{100}$--that is, $101^2=10,201$ Peres-Horodecki positive-partial-transpose tests were performed for {\it each} 
feasible sampled Tezuka-Faure point.
We present the two estimated bivariate separability functions in Fig.~\ref{fig:QubQut} and Fig.~\ref{fig:QubQut2} (cf. \cite[Figs. 3, 5]{slater833}).

\begin{figure}
\includegraphics{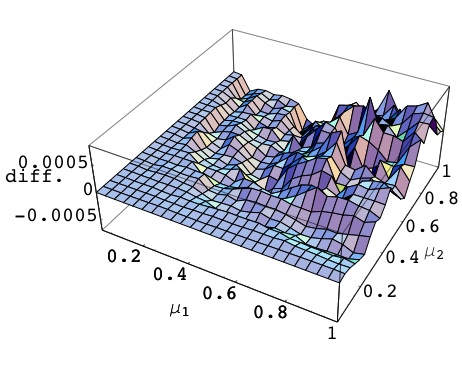}
\caption{\label{fig:Comparison}The complex 
qubit-qutrit separability function shown in Fig.~\ref{fig:QubQut2} minus
the {\it square} of the real qubit-qutrit separability 
function (Fig.~\ref{fig:QubQut2}), the latter function normalized so 
that the value in the plot at $\nu_1=1,\nu_2=1$ is 0. Note, importantly, 
the greatly-reduced $z$-axis scale {\it vis-{\`a}-vis} that of 
Fig.~\ref{fig:QubQut2}}
\end{figure}

In Fig.~\ref{fig:Comparison} we present a test of our Dyson-index
HS 
separability-function ansatz by subtracting from Fig.~\ref{fig:QubQut2}
the {\it square} of the function in Fig.~\ref{fig:QubQut}, 
which has been normalized
so that its value at $\nu_{1}=1,\nu_{2}=1$ equals that of the raw, 
unadjusted complex separability function.
One should, of course, note the greatly-reduced $z$-axis scale from
Fig.~\ref{fig:QubQut2}, indicating close adherence to the HS Dyson-index
separability-function ansatz, which it has been a principal goal
of this study to test.

\begin{figure}
\includegraphics{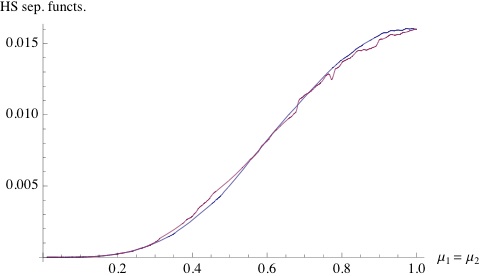}
\caption{\label{fig:Transect}The complex qubit-qutrit separability function
(Fig.~\ref{fig:QubQut2}) and the normalized square of the real function
(Fig.~\ref{fig:QubQut}), holding $\nu_1=\nu_2$. By construction, the two
curves are equal at $\nu_1=\nu_2=1$. The 
observed closeness of the two curves 
would be suggested by the HS Dyson-index ansatz}
\end{figure}

Now, in Fig.~\ref{fig:Transect} we plot two very closely-fitting curves.
One is the {\it complex} 
separability function holding $\nu_1=\nu_2$, and the other,
the {\it square} of the 
{\it real} separability function also holding $\nu_1=\nu_2$, but normalized
to equal the first (complex) function at the point 
$(1,1)$. This is also compelling
evidence for the validity of the HS Dyson-index ansatz.

\begin{figure}
\includegraphics{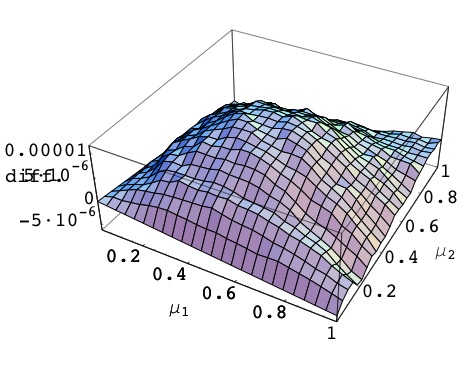}
\caption{\label{fig:NewFit}
The estimated qubit-qutrit real separability function (Fig.~\ref{fig:QubQut})
minus the candidate function (\ref{newcandidate}), the latter function being
scaled so the plotted value at $(\nu_1,\nu_2)=(1,1)$ is 0.}
\end{figure}

\begin{figure}
\includegraphics{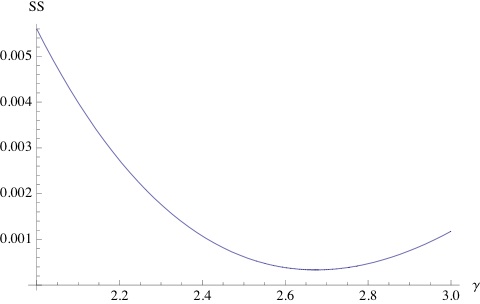}
\caption{\label{fig:LS1}The sum-of-squares (SS) of the fit of the
one-parameter family of functions $1-(1- \nu_1 \nu_2)^{\gamma}$ to the
normalized estimated real qubit-qutrit separability function
(Fig.~\ref{fig:QubQut}). For $\gamma=2.5$, we obtain the 
candidate separability function (\ref{newcandidate})}
\end{figure}

Also, in \cite{slater833}, we indicated that it strongly appeared that though
{\it two} ratio variables, $\nu_1$ and $\nu_2$, 
given in (\ref{tworatios}), are {\it ab initio} 
necessary in the qubit-qutrit analysis, it seems that upon further analysis 
they coalesce
into a  product 
\begin{equation}
\eta =\nu_1 \nu_2 = \frac{\rho_{11} \rho_{66}}{\rho_{33} \rho_{44}}, 
\end{equation}
and the 
separability function problem becomes actually simply univariate
in nature, rather than bivariate. This aspect needs, of course, to be more
closely evaluated in light of our new numerical results. 
In fact, one candidate HS separability function of such a {\it univariate}
nature which can be seen to fit 
our estimated functions (Figs.~\ref{fig:QubQut} 
and \ref{fig:QubQut2})
both very well (when appropriately normalized and/or squared)  is 
(Fig.~\ref{fig:NewFit})
\begin{equation} \label{newcandidate}
\mathcal{S}^{HS}_{real/qub-qut}(\nu_1 \nu_2) \propto 
1-\left(1-\nu _1 \nu _2\right)^{\frac{5}{2}} = 1-(1-\eta)^{\frac{5}{2}} 
= \frac{5}{2} B_{\eta}(1,\frac{5}{2}).
\end{equation}
In Fig.~\ref{fig:NewFit} we show the fit of this function to the 
estimated qubit-qutrit real separability function (Fig.~\ref{fig:QubQut}).
In Fig.~\ref{fig:LS1} we show the sum-of-squares of the fit of the 
one-parameter family of functions $1-(1- \nu_1 \nu_2)^{\gamma}$ to the 
normalized estimated real qubit-qutrit separability function 
(Fig.~\ref{fig:QubQut}). 
(For $\gamma=\frac{5}{2}$ we obtain (\ref{newcandidate}). We observe that
the minimum of the curve is somewhat in the neighborhood of 
$\gamma=\frac{5}{2}$.)

\begin{figure}
\includegraphics{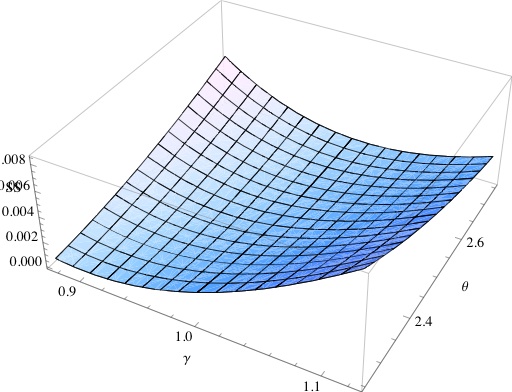}
\caption{\label{fig:LS2}The sum-of-squares (SS) of the fit of the
{\it two}-parameter family of functions 
$1-(1- (\nu_1 \nu_2)^{\theta})^{\gamma}$ to the
normalized estimated real qubit-qutrit separability function
(Fig.~\ref{fig:QubQut}). For $\gamma=2.5,\theta=1$, we obtain the 
candidate separability function (\ref{newcandidate})}
\end{figure}

In Fig.~\ref{fig:LS2} we show the sum-of-squares of the fit of the 
{\it two}-parameter family of functions 
$1-(1- (\nu_1 \nu_2)^{\theta})^{\gamma}$ to the 
normalized estimated real qubit-qutrit separability function 
(Fig.~\ref{fig:QubQut}). 
(For $\gamma=\frac{5}{2},\theta=1$ 
we obtain (\ref{newcandidate}). We observe that
the minimum of the curve is somewhat in the neighborhood of 
$\gamma=\frac{5}{2},\theta=1$.)

In the framework of Andai \cite{andai}, the total volume of
the 25-dimensional convex set of real $6 \times 6$ density matrices
can be represented as the product
\begin{equation}
V^{HS}_{qub-qut/real} = \frac{8192 \pi ^6}{253125}
\cdot \frac{25 \pi ^3}{1399771004732964864} 
= \frac{\pi ^9}{1730063650258944000}
\approx 1.72301 \cdot 10^{-14}
\end{equation}
and the total volume of
the 35-dimensional convex set of complex $6 \times 6$ density matrices 
as the product
\begin{equation}
V^{HS}_{qub-qut/complex} =  \frac{\pi ^{15}}{86400000} \cdot
 \frac{1}{3460550346681745424512204800} 
\end{equation}
\begin{displaymath}
= \frac{\pi ^{15}}{298991549953302804677854494720000000} 
\approx 9.58494 \cdot 10^{-29}.
\end{displaymath}
In both of these products, the first (20- or 30-dimensional) 
factor serves as the denominator
of the ratio $R_{1}$ and the second (5-dimensional) 
factor, as the denominator of the 
ratio $R_{2}$. (The numerator of $R_{1}$ is the 20- or 30-dimensional mass
assigned to the {\it separable} density matrices, and the numerator
of $R_{2}$ is the integral of the product 
of the corresponding suitably normalized separability
function and Bloore jacobian over the 5-dimensional unit simplex.)

Under the assumption of the correctness of (\ref{newcandidate}),
we find that the qubit-qutrit counterparts of the important constants
$R_{2}$ employed above in the two-qubit case ((\ref{ratio2}) and 
immediately below there) are 
\begin{equation}
R_{2_{qub-qut}}(\beta=1) = 1-\frac{4194304}{4849845 \pi }
\approx 0.724715,
\end{equation}
\begin{equation}
R_{2_{qub-qut}}(\beta=2) = \frac{-44632342463+68578836480 \log (2)}{4190140110}
\approx 0.692789,
\end{equation}
\begin{equation}
R_{2_{qub-qut}}(\beta=4) = 
\end{equation}
\begin{displaymath}
\frac{192210846322598002116984324520591-277301145703236210250598232096768
   \log (2)}{501570554133080277487570824}
\end{displaymath}
\begin{displaymath}
\approx 0.675902,
\end{displaymath}
for the real, complex and quaternionic cases, respectively. Again, we
note a monotonic decrease as the Dyson index $\beta$ increases.
Also, for the truncated quaternionic ($\beta=3$) case, we found
\begin{equation}
R_{2_{qub-qut}}(\beta=3) =
\end{equation}
\begin{displaymath}
-\frac{967504709}{552123}-\frac{18446744073709551616
   (-67294453713397888+5638997741091 \pi
   )}{71729672378917671400466262753675 \pi ^2}
\end{displaymath}
\begin{displaymath}
\approx  0.681261.
\end{displaymath}

Our sample estimates of the complementary $R_{1}$ constants were 0.226468, 
in the real case, and 0.047679 in the complex case. 
(We note that these estimates should be 
{\it independent} of the choice 
(correctness) of separability functions (\ref{newcandidate})
and, in the complex case, the square of (\ref{newcandidate}).) 
Forming the
products $R_{1} R_{2}$ based on these estimates, we obtain an
estimated real separability probability of 0.164125 and complex separability
probability of 0.0330446.

Our numerical analyses have been concerned only with pairs of values 
of $\nu_1$ and $\nu_2$ lying within the (bounded) unit square 
$[0,1] \times [0,1]$. We have made the implicit assumption that
the substance of our analyses would not be altered/biased if we were able to
incorporate all possible pairs lying within the {\it unbounded} quadrant
$[0,\infty] \times [0,\infty]$. (Clearly, for points $\nu_1>1,\nu_2>1$, 
simply 
by symmetry considerations, we immediately 
expect the separability function to be
proportional to $1-(1-\frac{1}{\eta})^{\frac{5}{2}}$. For pairs of points 
$(\nu_1,\nu_2)$ 
for which one member is greater than 1 and the other less than one, 
we expect the form the separability function takes to depend on whether 
or not $\eta= \nu_1 \nu_2 >1$.)
\subsubsection{Relations to previous qubit-qutrit analyses 
\cite{slaterPRA2,slater833}}
In \cite{slaterPRA}, we 
had undertaken 
a large-scale numerical 
(again, Tezuka-Faure quasi-Monte Carlo 
integration) analysis of the separable volumes of 
the 35-dimensional convex set of {\it complex} qubit-qutrit systems, endowed with the Hilbert-Schmidt, as well as a number of monotone (including the 
Bures) metrics. The estimate we obtained there of the HS separability
probability was 0.0266891. As we pointed out in our subsequent study
\cite[p. 14305]{slater833}, this is remarkably close to 
$\frac{32}{1199} \approx 0.0266889$.

In the subsequent study 
\cite{slater833}, which was chiefly devoted to the case of
two-qubit systems, we had included 
supplementary analyses of the real and complex
qubit-qutrit systems. But there we had only employed--in the interest of 
alacrity--Monte Carlo (random number), rather than 
(``lower-discrepancy'') 
quasi-Monte Carlo methods. So, the results presented in this paper, 
we believe
should be more accurate and informative. (Also, rather than sampling grids 
for $\nu_1,\nu_2$ of size $101 \times 101$, we had employed
grids of sizes $50 \times 50$ in the real case, and $20 \times 20$ in the complex case.) In \cite[sec.~10]{slater833}, we had 
put forth the tentative hypothesis that the corresponding real
qubit-qutrit separability function was simply proportional to 
$\sqrt{\eta} = \sqrt{\nu_1 \nu_2}$ (having a somewhat similar profile to 
our present candidate 
(\ref{newcandidate}), but definitely providing an inferior fit to our 
numerical results here). 
\subsection{Bures two-qubit separability functions and probabilities} \label{secBures}
Now, in our present study, 
we shall somewhat parallel the sequential approach of
\.Zyczkowski and Sommers in that they, first,  computed the {\it 
total} volume of
(separable {\it and} nonseparable) $n \times n$ density matrices in terms of
the (flat or Euclidean) 
Hilbert-Schmidt metric 
\cite{szHS} \cite[secs. 9.6-9.6, 14.3]{ingemarkarol}, and then,
using the 
fundamentally important 
Bures (minimal monotone) metric \cite[sec. 14.4]{ingemarkarol} 
 \cite{szBures}.
(In particular, they employed the Laguerre ensemble of random matrix theory 
\cite{random} 
in both sets of computations (cf. \cite{andai}). The Bures and HS metrics
were compared by Hall \cite{hall}, who concluded that the Bures induced
the ``minimal-knowledge ensemble'' (cf. \cite{slatersrednicki}), also 
noting that in the single-qubit case, the Bures metric ``may be recognized 
as the spatial part of the Robertson-Walker metric in general relativity''.)
That is, we will seek now
to extend the form of  analysis applied in the Hilbert-Schmidt context in
\cite{slater833} to the Bures setting.
\subsubsection{Review of earlier parallel Hilbert-Schmidt findings}
To begin, let us review the most elementary findings reported in 
\cite[sec. II.A.1]{slater833}. 
The simplest (four-parameter) scenario studied there posits a 
$4 \times 4$ density matrix $\rho$ with
fully general diagonal entries ($\rho_{11}, \rho_{22}, \rho_{33}, 
\rho_{44} = 1 -\rho_{11}-\rho_{22}-\rho_{33}$) and only one pair of real off-diagonal non-zero entries,
$\rho_{23}=\rho_{32}$. The HS separability function 
for that scenario was found to take 
the form
\cite[eq. (20)]{slater833},
\begin{equation} \label{equationA}
\mathcal{S}^{HS}_{[(2,3)]}(\mu) =
\begin{cases}
 2 \mu & 0\leq \mu \leq 1 \\
 2 & \mu >1
\end{cases},
\end{equation}
where we primarily employ
the variable $\mu=
\sqrt{\frac{\rho_{11} \rho_{44}}{\rho_{22} \rho_{33}}}$, rather than 
$\nu=\mu^2$, as in \cite{slater833,slaterPRA2}.

Allowing the 23- and 32-entries to be complex conjugates of one another,
we further found for the corresponding 
separability function \cite[eq. (22)]{slater833}---where the 
wide tilde over an ${i,j}$ pair will 
throughout indicate a complex entry (described by  
{\it two} parameters)---
\begin{equation} \label{equationB}
\mathcal{S}^{HS}_{[\widetilde{(2,3)}]}(\mu)=
(\sqrt{\frac{\pi}{2}} \mathcal{S}^{HS}_{[(2,3)]}(\mu))^2 =
\begin{cases}
 \pi  \mu^2  & 0\leq \mu \leq 1 \\
 \pi  & \mu >1
\end{cases}.
\end{equation}

Further, permitting 
the 23- and 32-entries to be {\it quaternionic}  conjugates of one another 
\cite{adler,asorey}, the corresponding
separability function \cite[eq. (24)]{slater833}---where the
wide hat over an ${i,j}$ pair will throughout indicate a quaternionic 
entry (described by
{\it four} parameters)---took the form
\begin{equation} \label{equationQuat}
\mathcal{S}^{HS}_{[\widehat{(2,3)}]}(\mu)= 
(\sqrt{\frac{\pi}{2}} \mathcal{S}^{HS}_{[\widetilde{(2,3)}]}(\mu))^2 =
(\sqrt{\frac{\pi}{2}} \mathcal{S}^{HS}_{[(2,3)]}(\mu))^4=
\begin{cases}
 \frac{\pi^2  \mu^4}{2}  & 0\leq \mu \leq 1 \\
 \frac{\pi^2}{2}  & \mu >1
\end{cases}.
\end{equation}

So, the real (\ref{equationA}), complex (\ref{equationB}), 
and quaternionic (\ref{equationQuat})
HS separability functions accord {\it perfectly}
with the Dyson index sequence  $\beta= 1, 2, 4 $ of random matrix theory
\cite{dyson}. ``The value of $\beta$ is given by the number of independent
degrees of freedom per matrix element and is determined by the antiunitary 
symmetries \ldots It is a concept that originated in Random Matrix Theory 
and is important for the Cartan classification of symmetric spaces'' 
\cite[p. 480]{kogut}. The Dyson index corresponds to the ``multiplicity of 
ordinary roots'', in the terminology of symmetric spaces \cite[Table 2]{caselle}. 
However, we remain unaware of any specific line of argument using random
matrix theory \cite{random} that can be used to formally confirm
the HS separability function Dyson-index-sequence 
phenomena we have noted above and observed in 
\cite{slater833}. (The basic difficulty/novelty 
appears to be that the separability
aspect of the problem introduces a totally new set of complicated
constraints---{\it quartic} (biquadratic) in $\mu$ 
\cite[eq. (5)]{slater833} \cite[eq. (7)]{slaterPRA2}---that the multivariate integration must respect 
\cite[sec. I.C]{slater833}.)

As a new exercise here, unreported in \cite{slater833}, 
we found that
setting any single one of the four components of the quaternionic 
entry, $x_{23} +{\bf{i}}  y_{23} +{\bf{j}} j u_{23} +{\bf{k}} 
v_{23}$, in the
scenario just described, to zero, yields the 
(truncated quaternionic) separability function,
\begin{equation} \label{missing1}
\mathcal{S}^{HS}_{[\hat{(2,3)}]} =
\begin{cases}
 \frac{4 \pi  \mu ^3}{3} & 0\leq \mu \leq 1 \\
 \frac{4 \pi }{3} & \mu >1
\end{cases},
\end{equation}
consistent, at least, in terms 
of the exponent of $\mu$, with the Dyson-index 
pattern previously observed.

Continuing the analysis in \cite{slater833}, we computed 
the integrals 
\begin{equation} \label{Vsmall}
V^{HS}_{sep/scenario}= \int_{0}^{\infty} \mathcal{S}^{HS}_{scenario}(\mu) 
\mathcal{J}^{HS}_{scenario}(\mu)
d \mu,
\end{equation}
of the products of 
these separability functions with the corresponding (univariate) 
marginal jacobian functions
 (which are obtained by integration over
diagonal parameters only  and {\it not} any of 
the off-diagonal $x_{ij}$'s and 
$y_{ij}$'s) for the reparameterization of $\rho$ using the Bloore variables
\cite[eq. (17)]{slater833}. This 
yielded the HS scenario-specific {\it separable} volumes 
$V^{HS}_{sep/scenario}$. The ratios of such separable volumes  to
the HS total volumes
\begin{equation} \label{Vbig}
V^{HS}_{tot/scenario}= c_{scenario}^{HS} 
\int_{0}^{\infty}  \mathcal{J}_{scenario}^{HS}(\mu) d \mu,
\end{equation}
where $c^{HS}_{scenario}$ is a scenario-specific constant, gave 
us in \cite{slater833} (invariably, it seems, exact) separability {\it probabilities}. (For the three scenarios listed above, these probabilities were, 
respectively, $\frac{3 \pi}{16}, \frac{1}{3}$ and $\frac{1}{10}$.)

Based on the numerous scenario-specific 
analyses in \cite{slater833}, we are led to believe
that the real, complex and quaternionic separability functions conform to
the Dyson-index pattern for general scenarios, when
the Hilbert-Schmidt measure has been employed. This apparent adherence
was of central importance 
in arriving at the assertions in \cite[secs.~IX.A.1 and 
IX.A.2]{slater833} that the HS separability probabilities of generic 
(9-dimensional] real)
and (15-dimensional) 
complex two-qubit states 
are $\frac{8}{17}$ and 
$\frac{8}{33}$, respectively. 
There we had posited---using mutually supporting numerical and theoretical 
arguments---that \cite[eq. (102)]{slater833}
\begin{equation}
\mathcal{S}^{HS}_{real}(\mu) \propto  \frac{1}{2} (3- \mu^2) \mu,
\end{equation}
and, further pursuing our basic Dyson-index ansatz (fitting our 
numerical simulation extremely well \cite[Fig. 4]{slater833}), that
$(\mathcal{S}^{HS}_{real}(\mu))^2 \propto \mathcal{S}^{HS}_{complex}(\mu)$.
(Also, in the first part of the analyses above, we presented numerical 
evidence 
 that $(\mathcal{S}^{HS}_{real}(\mu))^4 
\propto \mathcal{S}^{HS}_{quat}(\mu)$, and made this relation
more precise (\ref{bigfish}).)
\subsubsection{Four-parameter density-matrix scenarios}
Now, employing 
formulas (13) and (14) 
of Dittmann \cite{explicit} for the {\it Bures} metric---which 
avoid the possibly problematical need for diagonalization 
of $\rho$---we were able to find the
Bures volume elements for the same three 
basic (one pair of free off-diagonal entries) 
scenarios. We obtained for the real case,
\begin{equation} \label{V23real}
dV^{Bures}_{[(2,3)]} = \frac{\sqrt{\rho _{11}} \sqrt{1-\rho _{11}-\rho _{22}}
   \sqrt{\rho _{22}}}{4 \sqrt{1-x_{23}^2} \left(\rho _{22}
   \mu ^2+\rho _{11}\right) \sqrt{\mu ^2 \rho
   _{22}^2+\left(1-\rho _{11}\right) \rho _{11}}} d \rho_{11} d\rho_{22} d x_{23} d \mu,
\end{equation}
for the complex case,
\begin{equation} \label{V23complex}
dV^{Bures}_{[\widetilde{(2,3)}]}= \frac{\rho _{11} \rho _{22} \left(\rho _{11}+\rho
   _{22}-1\right)}{4 \sqrt{1-y_{23}^2-x_{23}^2}
   \left(\rho _{22} \mu ^2+\rho _{11}\right) \left(-\rho
   _{11}^2+\rho _{11}+\mu ^2 \rho _{22}^2\right)} 
d \rho_{11} d\rho_{22} d x_{23} d y_{23} d \mu,
\end{equation}
and for the quaternionic case,
\begin{equation} \label{V23quaternionic}
dV^{Bures}_{[\widehat{(2,3)}]}= \frac{A}{B} 
d \rho_{11} d\rho_{22} d x_{23} d y_{23}  d u_{23} d v_{23} d \mu,
\end{equation}
where
\begin{displaymath}
A=  -\rho _{11}^2 \rho _{22}^2 \left(\rho _{11}+\rho
   _{22}-1\right)^2,
\end{displaymath}
and
\begin{displaymath}
B=4 \sqrt{1-u_{23}^2-v_{23}^2-x_{23}^2-y_{23}^2} \left(\rho
   _{22} \mu ^2+\rho _{11}\right) \left(-\rho _{11}^2+\rho
   _{11}+\mu ^2 \rho _{22}^2\right)^2.
\end{displaymath}

In analyzing the
quaternionic case, we transformed---using standard 
procedures \cite[p. 495]{adler} \cite[eq. (17)]{slaterJMP1996}---the 
corresponding $4 \times 4$ density matrix into an $8 \times 8$ density matrix with (only) complex entries. To
this, we found it most convenient to apply---since its 
eigenvalues and eigenvectors 
could be explicitly computed---the basic formula of H\"ubner 
\cite{hubner} \cite[p. 2664]{dittmann} for the Bures metric. 

Integrating these three volume elements  over all the 
(four, five or seven) 
variables, while enforcing the nonnegative definiteness 
requirement for $\rho$, we derived
the Bures 
{\it total} (separable {\it and} nonseparable) volumes for the three
scenarios---$V^{Bures}_{tot/[(2,3)]} =\frac{\pi^2}{12} 
\approx 0.822467$, 
$V^{Bures}_{tot/[\widetilde{(2,3)}]}=\frac{\pi^3}{64} \approx 0.484473$, and 
$V^{Bures}_{tot/[\widehat{(2,3)}]}=\frac{\pi^4}{768} \approx 0.126835$.

We note importantly that
the Bures volume elements ((\ref{V23real}), (\ref{V23complex}), 
(\ref{V23quaternionic})), in these three cases, can be 
{\it factored} into products of functions of 
the {\it off-diagonal} Bloore
 variables, $u_{23}, v_{23}, x_{23}$ and $y_{23}$, 
and functions of the {\it diagonal} 
variables, $\rho_{11}, \rho_{22}$ and $\mu$. Now, we will integrate 
(one may transform to polar and spherical coordinates, as appropriate) 
just the 
factors ---$\frac{1}{\sqrt{1-x_{23}^2}}$, 
$\frac{1}{\sqrt{1-x_{23}^2 - y_{23}^2}}$ 
and $\frac{1}{\sqrt{1-u_{23}^3-v_{23}^2-x_{23}^2-y_{23}^2}}$---involving
the off-diagonal variable(s) over 
those variables. In doing this, we will further enforce 
(using the recently-incorporated 
integration-over-implicitly-defined-regions feature of Mathematica)
the Peres-Horodecki positive-partial-transpose-criterion
\cite{asher,michal,bruss}, 
expressible as
\begin{equation}
\mu^2 -x_{23}^2 \geq 0
\end{equation}
 in the real case,
\begin{equation}
\mu^2 -x_{23}^2 -y_{23}^2 \geq 0,
\end{equation}
 in the complex case, and 
\begin{equation}
\mu^2 -x_{23}^2 -y_{23}^2 - u_{23}^2 -v_{23}^2 \geq 0,
\end{equation}
in the quaternionic case. (None of the individual diagonal $\rho_{ii}$'s 
appears explicitly in these constraints, due to an attractive property of the
Bloore [correlation coefficient/off-diagonal scaling] 
parameterization. Replacing $\mu^2$ in these three constraints by 
simply unity, we obtain the non-negative definiteness constraints on $\rho$ itself, which we also obviously must enforce.) 
Performing the indicated three integrations, we obtain
the {\it Bures} separability functions,
\begin{equation} \label{Bures1}
\mathcal{S}^{Bures}_{[(2,3)]}(\mu) = 
\begin{cases}
 \pi  & \mu \geq 1 \\
 2 \sin ^{-1}(\mu ) & 0 < \mu <1
\end{cases},
\end{equation}
\begin{equation} \label{Bures2}
\mathcal{S}^{Bures}_{[\widetilde{(2,3)}]}(\mu) = 
\begin{cases}
 2 \pi  & \mu \geq 1 \\
 2 \pi  \left(1-\sqrt{1-\mu ^2}\right) & 0<\mu <1
\end{cases},
\end{equation}
and
\begin{equation} \label{Bures3}
\mathcal{S}^{Bures}_{[\widehat{(2,3)}]}(\mu) =
\begin{cases}
 \frac{4 \pi ^2}{3} & \mu >1 \\
 \frac{2}{3} \pi ^2 \left(-\sqrt{1-\mu ^2} \mu ^2-2
   \sqrt{1-\mu ^2}+2\right) & 0 <\mu <1
\end{cases}.
\end{equation}
Then, utilizing these three separability functions---that is, 
integrating the products of the functions and the 
corresponding remaining 
{\it diagonal}-variable factors 
in the Bures volume elements ((\ref{V23real}), (\ref{V23complex})), 
((\ref{V23quaternionic})) 
over the $\mu, \rho_{11}$ and $\rho_{22}$ variables---we obtain
{\it separable} volumes of $V^{Bures}_{sep/[(2,3)]}= 0.3658435525$ and
\begin{equation}
V^{Bures}_{sep/[\widetilde{(2,3)}]}=  
V^{Bures}_{tot/[\widetilde{(2,3)}]} - \frac{1}{32} \pi ^2 (-2 C+\pi ) = 
\frac{1}{64} \pi ^2 (4 C-6 +\pi ) 
\approx 0.124211 
\end{equation}
 and consequent
separability {\it probabilities}, respectively, 
of 0.4448124200 and (our only {\it exact} Bures separability probability
result in this study (cf. \cite{slaterC})),
\begin{equation} \label{exactprob}
P^{Bures}_{sep/[\widetilde{(2,3)}]} = \frac{4 C-6+\pi }{\pi } 
\approx 0.256384,
\end{equation}
where $C \approx 0.915966$ is Catalan's constant (cf. \cite{collins}). 
(This constant appears
commonly in estimates of 
combinatorial functions and in certain classes of sums and definite 
integrals \cite[sec.~1.7]{finch}. 
The ratio $\frac{C}{\pi}$--as well as
having an interesting series expansion \cite[p. 54]{finch}--occurs in
exact solutions to the dimer problem of
statistical mechanics  \cite[p. 54]{finch} \cite{temperley,gagun}).

Further, for the quaternionic case, 
$V^{Bures}_{sep/\widehat{[(2,3)]}} \approx 0.012954754466$, and 
$P^{Bures}_{sep/\widehat{[(2,3)]}} \approx  0.10213883862$.
 (The corresponding HS separability probability was also of the 
same relatively 
small magnitude, that is, $\frac{1}{10}$ \cite[sec.~II.A.3]{slater833}.
We have computed the various Bures separable volumes and probabilities
to high numerical accuracy, hoping that such accuracy may be useful
in searches for possible further exact formulas for them.)

\begin{figure}
\includegraphics{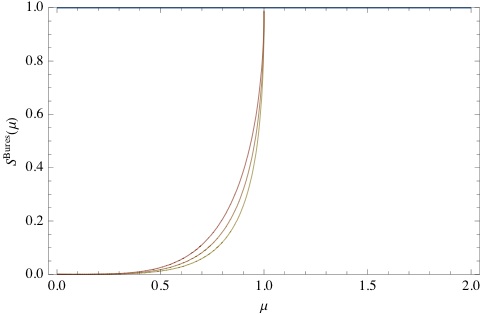}
\caption{\label{fig:functs} Joint plot of the 
normalized Bures {\it quaternionic} 
separability function 
$\frac{3 \mathcal{S}^{Bures}_{[\widehat{(2,3)}]}(\mu)}{4 \pi^2}$, 
the {\it square} of 
the normalized Bures {\it complex} separability function
$\frac{\mathcal{S}^{Bures}_{[\widetilde{(2,3)}]}(\mu)}{2 \pi}$,
and the {\it fourth} power of the normalized Bures {\it real} separability
function $\frac{\mathcal{S}^{Bures}_{[(2,3)]}(\mu)}{\pi}$. The order of
dominance of the three curves is the same as the order in which they have been
mentioned.}
\end{figure}

So, the normalized---to equal 1 at $\mu=1$---forms of these three
separability 
functions are $\frac{\mathcal{S}^{Bures}_{[(2,3)]}(\mu)}{\pi}$,
$\frac{\mathcal{S}^{Bures}_{[\widetilde{(2,3)}]}(\mu)}{2 \pi}$ and 
$\frac{3 \mathcal{S}^{Bures}_{[\widehat{(2,3)}]}(\mu)}{4 \pi^2}$. 
In Fig.~\ref{fig:functs}, we plot---motivated by the appearance of the 
Dyson indices in the analyses of 
\cite{slater833}---the {\it fourth} power of the first (real) of these
three normalized functions together with the {\it square} of the 
second (complex) function and the (untransformed) third 
(quaternionic) function itself. 
We find a very close, 
\begin{equation}
\Big(\frac{\mathcal{S}^{Bures}_{[(2,3)]}(\mu)}{\pi}\Big)^4  \approx (\frac{\mathcal{S}^{Bures}_{[\widetilde{(2,3)}]}(\mu)}{2 \pi})^2 
\approx (\frac{3 \mathcal{S}^{Bures}_{[\widehat{(2,3)}]}(\mu)}{4 \pi^2}),
\end{equation}
but now {\it not} exact fit, as we did find 
in \cite{slater833}  for their (also normalized) 
Hilbert-Schmidt
counterparts  $\frac{\mathcal{S}^{HS}_{[(2,3)]}(\mu)}{2}$,
$\frac{\mathcal{S}^{HS}_{[\widetilde{(2,3)}]}(\mu)}{\pi}$ 
and $\frac{2 \mathcal{S}^{HS}_{[\widehat{(2,3)}]}(\mu)}{\pi^2}$ ((\ref{equationA}), (\ref{equationB}), (\ref{equationQuat})).

As an additional exercise (cf. (\ref{missing1})), 
we have computed the Bures separability function
in the (truncated quaternionic) 
case that a single one of the four components of 
the (2,3)-quaternionic entry is set to zero.
Then, we have (falling into the same tight cluster in Fig.~\ref{fig:functs},
when the $\frac{4}{3}$-power of its
normalized form is plotted)
\begin{equation} \label{missing2}
\mathcal{S}^{Bures}_{[\hat{(2,3)}]}=
\begin{cases}
 \frac{1}{8} \pi ^2 \left(4-\sqrt{2} \log \left(3+2
   \sqrt{2}\right)\right) & \mu >1 \\
 \frac{1}{4} \pi  \left(\mu  \sqrt{1-\mu ^2}-\sin
   ^{-1}(\mu )\right) \left(\sqrt{2} \log \left(3+2
   \sqrt{2}-4\right)\right) & 0 < \mu <1
\end{cases}.
\end{equation}

We have been able, further, using the formulas of Dittmann \cite{explicit},
to compute the Bures volume elements for the 
corresponding (five-dimensional) real and 
(seven-dimensional)  complex scenarios, in which {\it both} the $\{2,3\}$ and $\{1,2\}$ entries are allowed to 
freely vary.  But these volume elements do not appear, now, 
to fully factorize into
products of functions 
(as is the case for (\ref{V23real}) and (\ref{V23complex})) 
involving just $\rho_{11}, \rho_{22}, \mu$  and just the off-diagonal
variables $x_{ij}$'s and $y_{ij}$'s. The requisite integrations are, then, 
more problematical and it seemed impossible to obtain an 
explicit univariate
separability function of $\mu$.

For instance, in this regard, we have for the indicated five-dimensional real 
scenario that 
\begin{equation}
dV^{Bures}_{[(1,2),(2,3)]}=
\frac{1}{4} \sqrt{\frac{A}{B C (D +E)}} d \rho_{11} d\rho_{22} 
d x_{12} d x_{23}  d \mu,
\end{equation}
where
\begin{equation}
A=-\rho _{11}^2 \rho _{22}^2 \left(\rho _{11}+\rho
   _{22}-1\right) \left(\left(\mu ^2-1\right) \rho
   _{22}+1\right),
\end{equation}
\begin{equation}
B= \left(\rho _{22} \mu ^2+\rho _{11}\right)^2, C=x_{12}^2+x_{23}^2-1,
\end{equation}
\begin{equation}
D= \left(\rho _{11}+\rho _{22}\right) \left(x_{12}^2 \rho
   _{22} \left(\rho _{22} \mu ^2+\rho
   _{11}\right)^2-\left(\left(\mu ^2-1\right) \rho
   _{22}+1\right) \left(-\rho _{11}^2+\rho _{11}+\mu ^2
   \rho _{22}^2\right)\right)
\end{equation}
and
\begin{equation}
E=-x_{23}^2 \rho _{22} \left(\rho _{11}+\rho _{22}-1\right)
   \left(-\rho _{11}^2+\rho _{11}+\mu ^2 \rho
   _{22}^2\right).
\end{equation}
So, no desired factorization is apparent.
\subsubsection{Five-parameter density-matrix scenarios}
However, the computational situation greatly improves if we let the (1,4) 
and (2,3)-entries be the two free ones. (These entries are the specific ones
that are interchanged under 
the operation of partial transposition, so there is a greater 
evident symmetry in such a scenario.) 
Then, we found that the three Bures 
volume elements all do factorize into products of functions of
off-diagonal entries and functions of diagonal entries. We have
\begin{equation}
dV^{Bures}_{[(1,4),(2,3)]} = 
\frac{1}{8} \sqrt{-\frac{1}{\left(x_{14}^2-1\right)
   \left(x_{23}^2-1\right) \left(\rho _{22}+\rho
   _{33}-1\right) \left(\rho _{22}+\rho _{33}\right)}} d \rho_{11} 
d \rho_{22} d \rho_{33} d x_{14} d x_{23},
\end{equation}
where simply for succinctness, we now show the volume elements before 
replacing the $\rho_{33}$ variable by $\mu$. 
(We note that the expression for 
$dV^{Bures}_{[(1,4),(2,3)]}$ is independent of 
$\rho_{11}$.)
For the corresponding complex scenario,
\begin{equation}
dV^{Bures}_{[\widetilde{(1,4)},\widetilde{(2,3)}]} =
\frac{1}{8} \sqrt{\frac{F}{G}}  d \rho_{11}
d \rho_{22} d \rho_{33} d r_{14} d r_{23} d \theta_{14} d \theta_{23},
\end{equation}
where
\begin{equation}
F=-r_{14}^2 r_{23}^2 \rho _{11} \rho _{22} \rho _{33}
   \left(\rho _{11}+\rho _{22}+\rho _{33}-1\right),
\end{equation}
and
\begin{equation}
G=\left(r_{14}^2-1\right) \left(r_{23}^2-1\right) \left(\rho
   _{22}+\rho _{33}-1\right)^2 \left(\rho _{22}+\rho
   _{33}\right)^2,
\end{equation}
and we have now further shifted to polar coordinates, 
$x_{ij} + {\bf{i}}  y_{ij} = r_{ij} (\cos{\theta_{ij}} + {\bf{i}} 
\sin{\theta_{ij}})$. 
For the quaternionic scenario, we have 
(using two sets of hyperspherical coordinates $(r_{14}, \theta_{14}^{(1)}, \theta_{14}^{(2)},\theta_{14}^{(3)})$ and $(r_{23}, \theta_{23}^{(1)}, \theta_{23}^{(2)}, \theta_{23}^{(3)})$),
\begin{equation}
dV^{Bures}_{[\widehat{(1,4)},\widehat{(2,3)}]} = \frac{1}{8} \sqrt{\frac{\tilde{F}}{\tilde{G}}}  d \rho_{11}
d \rho_{22} d \rho_{33} d r_{14} d r_{23} d \theta_{14}^{(1)} d \theta_{14}^{(2)} d \theta_{14}^{(3)} d \theta_{23}^{(1)} d \theta_{23}^{(2)} d \theta_{23}^{(3)},
\end{equation}
where
\begin{equation}
\tilde{F}=\sin ^2\left(\theta _{14}^{(1)}\right) \sin \left(\theta _{14}^{(2)}\right)
   \sin ^2\left(\theta_{23}^{(1)}\right) \sin \left(\theta_{23}^{(2)}\right)
   r_{14}^3 r_{23}^3 \rho _{11}^{3/2} \rho _{22}^{3/2}
   \left(-\rho _{11}-\rho _{22}-\rho _{33}+1\right)^{3/2}
   \rho _{33}^{3/2}
\end{equation}
and
\begin{equation}
\tilde{G}=\sqrt{1-r_{14}^2} \sqrt{1-r_{23}^2} \left(\rho _{22}+\rho
   _{33}-1\right)^2 \left(\rho _{22}+\rho _{33}\right)^2.
\end{equation}

The total Bures volume for the first (real) of these three scenarios is 
$V^{Bures}_{tot/[(1,4),(2,3)]} = \frac{\pi^3}{64} \approx 0.484473$, for the second (complex) 
scenario, $V^{Bures}_{tot/[\widetilde{(1,4)},\widetilde{(2,3)}]} = 
\frac{\pi^4}{192} \approx 0.507339$, and for the third (quaternionic), 
$V^{Bures}_{tot/[\widehat{(1,4)},\widehat{(2,3)}]} = 
\frac{\pi^6}{245760} \approx 0.0039119$.

In the 
two corresponding 
Hilbert-Schmidt (real and complex) analyses 
we have previously reported, we had the results  \cite[eq. (28)]{slater833},
\begin{equation} \label{suggestion}
\mathcal{S}^{HS}_{[(1,4),(2,3)]}(\mu) = 
\begin{cases}
 4 \mu & 0\leq \mu \leq 1 \\
 \frac{4}{\mu} & \mu >1
\end{cases}.
\end{equation}
and \cite[eq. (34)]{slater833}
\begin{equation} \label{secondmixed}
\mathcal{S}^{HS}_{[\widetilde{(1,4)},\widetilde{(2,3)}]}(\mu)  =
\begin{cases}
 \pi ^2 \mu^2 & 0\leq \mu \leq 1 \\
 \frac{\pi ^2}{\mu^2 } & \mu >1
\end{cases},
\end{equation}
thus, exhibiting the indicated exact (Dyson-index sequence) 
proportionality relation. 
We now found, for the two Bures analogs,  that 
\begin{equation} \label{suggestion2}
\mathcal{S}^{Bures}_{[(1,4),(2,3)]}(\mu) =
\begin{cases}
 \pi ^2 & \mu =1 \\
 2 \pi  \csc ^{-1}(\mu ) & \mu >1 \\
 2 \pi  \sin ^{-1}(\mu ) & 0<\mu <1
\end{cases},
\end{equation}
\begin{equation} \label{secondmixed2}
\mathcal{S}^{Bures}_{[\widetilde{(1,4)},\widetilde{(2,3)}]}(\mu)  =
\begin{cases}
 16 \pi^2 & \mu =1 \\
 16 \pi^2 \left(1-\frac{\sqrt{\mu ^2-1}}{\mu }\right) & \mu >1
   \\
 16 \pi^2 \left(1-\sqrt{1-\mu ^2}\right) & 0<\mu <1
\end{cases},
\end{equation}
and, further still, for the quaternionic scenario,
\begin{equation} \label{thirdmixed2}
\mathcal{S}^{Bures}_{[\widehat{(1,4)},\widehat{(2,3)}]}(\mu)  =
\begin{cases}
 \frac{16 \pi ^4}{9} & \mu =1 \\
 -\frac{8 \pi ^4 \left(2 \left(\sqrt{\mu ^2-1}-\mu \right)
   \mu ^2+\sqrt{\mu ^2-1}\right)}{9 \mu ^3} & \mu >1 \\
 \frac{8}{9} \pi ^4 \left(-\sqrt{1-\mu ^2} \mu ^2-2
   \sqrt{1-\mu ^2}+2\right) & 0<\mu <1
\end{cases}.
\end{equation}
Employing these several results, we obtained that 
$V^{Bures}_{sep/[(1,4),(2,3)]} \approx 0.1473885131$,
$V^{Bures}_{sep/[\widetilde{(1,4)},\widetilde{(2,3)}]} 
\approx 0.096915844$, and
$V^{Bures}_{sep/[\widehat{(1,4)},\widehat{(2,3)}]}
\approx 0.000471134100$
giving us real, 
complex and quaternionic separability probabilities of 0.3042243652,
0.19102778 and 0.120436049.

We see that for values of $\mu \in [0,1]$, the {\it 
normalized} forms of these
three Bures separability functions are {\it identical} to the three
obtained above ((\ref{Bures1}), (\ref{Bures2}), (\ref{Bures3})) 
for the corresponding {\it single}-nonzero-entry scenarios. 
While those earlier functions were all constant for $\mu>1$, we now have
symmetrical behavior about $\mu=1$ in the form, $\mathcal{S}^{Bures}_{scenario}(\mu) =
\mathcal{S}^{Bures}_{scenario}(\frac{1}{\mu})$.

\begin{figure}
\includegraphics{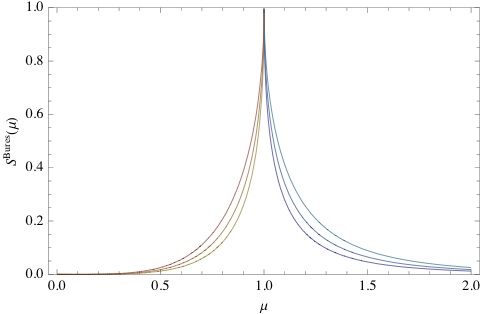}
\caption{\label{fig:functs2} Joint plot of the
normalized Bures {\it quaternionic}
separability function
$\frac{9 \mathcal{S}^{Bures}_{[\widehat{(1,4)},\widehat{(2,3)}]}(\mu)}{4}$,
the {\it square} of
the normalized Bures {\it complex} separability function
$\frac{\mathcal{S}^{Bures}_{[\widetilde{(1,4)},\widetilde{(2,3)}]}(\mu)}{16 \pi^2}$,
and the {\it fourth} power of the normalized Bures {\it real} separability
function $\frac{\mathcal{S}^{Bures}_{[(1,4),(2,3)]}(\mu)}{\pi^2}$. 
Over the interval
$\mu \in [0,1]$, the three functions are identical---with the same 
order of dominance---to those in 
Fig.~\ref{fig:functs}.}
\end{figure}

In Fig.~\ref{fig:functs2}, we show the analogous plot to Fig.~\ref{fig:functs},
using the normalized (to equal 1 at $\mu=1$) 
forms of the three  additional Bures separability functions 
((\ref{suggestion2}), (\ref{secondmixed2}), (\ref{thirdmixed2})). 
We again, of course, observe a very close fit to the type of proportionality relations
{\it exactly} observed in the Hilbert-Schmidt case 
((\ref{suggestion}), (\ref{secondmixed})).

We were, further, able to compute the Bures volume element for the
{\it three}-nonzero-entries
complex scenario $[\widetilde{(1,2)},\widetilde{(1,4)},
\widetilde{(2,3)}]$, but it was considerably more complicated in form than 
those reported above, so no 
additional analytical progress seemed possible.
\subsubsection{Additional analyses}
Regarding the possible computation of Bures separability functions for the 
{\it totality} of 
9-dimensional real and 15-dimensional complex two-qubit states, we have found,
preliminarily,  
that the corresponding metric tensors (using the Bloore parameterization 
[sec.~ \ref{secBloore}]) 
decompose into $3 \times 3$ and
$6 \times 6$, and $3 \times 3$ and $12 \times 12$ blocks, respectively.
The $3 \times 3$ blocks themselves 
are identical in the two cases, and of precisely 
the (simple diagonal) form (if we employ hyperspherical coordinates) that 
Akhtarshenas found for the Bures metric using the coset 
parameterization \cite[eq. (23)]{iran2}. These $3 \times 3$ 
blocks, thus, depend only upon the 
diagonal entries (while in \cite{iran2}, the dependence, quite 
differently, was upon the 
eigenvalues). It appears, 
though, that the determinants---for which we presently lack 
succinct formulas---of the complementary
$6 \times 6$ and $12 \times 12$ blocks, do depend upon all, diagonal and
non-diagonal, parameters, rendering further analytical progress, along 
the lines pursued with substantial success 
for the Hilbert-Schmidt metric, for these
scenarios rather problematical.
\subsubsection{Discussion}
The close proximity observed above between 
certain two-qubit separability results for the 
Hilbert-Schmidt and Bures metrics is perhaps somewhat similar in nature/explanation to a form of high similarity also observed in our previous analysis
\cite{slaterPRA}. There, large scale numerical (quasi-Monte Carlo) 
analyses strongly suggested that the ratio of Hilbert-Schmidt separability
probabilities of generic (rank-6) 
qubit-qutrit states ($6 \times 6$ density matrices) 
to the separability probabilities
of generically minimally degenerate (boundary/rank-5) 
qubit-qutrit states was equal to 2. (This has since been formally 
confirmed and generalized---in terms of 
positive-partial-transpose-ratios---to arbitrary bipartite systems
by Szarek, Bengtsson and 
{\.Z}yczkowski in \cite{sbz}. 
They found that the set of positive-partial-transpose states 
is ``pyramid decomposable'' and, hence, is a body of constant height.) 
Parallel numerical ratio estimates 
also obtained in \cite{slaterPRA} based on 
the Bures (and a number of other monotone) metrics were also surprisingly close to 2, as well (1.94334 in the Bures case \cite[Table IX]{slaterPRA}). 
However, no exact value for the Bures qubit-qutrit ratio has yet been 
established, and our separability function 
results above, might be taken to suggest that
the actual Bures ratio is not exactly equal to 2, but only quite close to it.
(Possibly, in these regards, 
the Bures metric might profitably be considered as some
perturbation of the flat Euclidean metric (cf. \cite{gross}).)

Further study of the forms the Bures separability functions take
for qubit-qubit and qubit-qutrit scenarios is, of course, possible, 
with the hope that one can gain
as much insight into the nature of Bures separability probabilities as 
has been obtained by examining the analogous
Hilbert-Schmidt separability functions \cite{slater833}.
(In \cite{slaterJGP}, we had formulated, based on extensive numerical
evidence, conjectures---involving the silver mean, $\sqrt{2}-1$---for the
Bures [and other monotone metric]
separability probabilities of the 15-dimensional convex set of
[complex] qubit-qubit states, which we would further aspire to test. 
One may also consider the use of monotone metrics other than the 
{\it minimal} Bures one 
\cite{andai}---such as the Kubo-Mori and Wigner-Yanase.)
The analytical challenges to further progress,
however, in light of the apparent non-factorizability of volume elements 
into diagonal and off-diagonal terms, 
appear quite formidable.
\section{Euler-angle-parameterization separability functions} \label{secEuler}
In the previous section (sec.~\ref{secBures}), 
we found that the Bloore parameterization (sec.~\ref{secBloore}), 
markedly
useful in determining separable volumes based on the (non-monotone) 
Hilbert-Schmidt metric, is less immediately fruitful when the Bures
(and presumably other monotone) metrics is employed.
In light of this development, 
it appeared to be of interest to see if some 
other parameterizations might prove 
amenable to such type of ``separability function'' analyses. 
In particular, we will examine here the use for such purposes--as 
earlier suggested 
in \cite{slaterJPAreject}--of the $SU(4)$ 
Euler-angle parameterization of the 15-dimensional 
complex set of $4 \times 4$ density
matrices developed by Tilma, Byrd and Sudarshan \cite{tbs}. 
(We will closely follow the notation and terminology of \cite{tbs}. In 
\cite{slaterJPAreject}, we simply attempted to fit symmetric 
polynomials \cite{ig} to yield previously-conjectured separable 
volumes, and did not initiate any 
independent quasi-Monte Carlo estimation procedures, as we will 
immediately below.)
\subsection{Complex two-qubit case}
The fifteen parameters, then, 
employed will be twelve Euler angles ($\alpha_{i}$'s)
and three independent eigenvalues ($\lambda_{1}, \lambda_2, \lambda_3$). 
The {\it total} (separable {\it and} nonseparable) volume is simply
(for all metrics of interest) 
the {\it product} of integrals over these two sets of parameters 
\cite{szHS,szBures}.
Now, to study the separable-volume question, 
in complete analogy to our methodology
above, we would like to integrate over the twelve Euler angles 
(rather than the off-diagonal Bloore parameters, as before), while
enforcing the positivity of the partial transpose, required for 
separability. We were, fortunately, able to perform such enforcement 
in the Bloore-parameterization case employing 
only a {\it single} diagonally-related 
parameter ($\mu$), given in (\ref{firstratio}). Such a 
reduction in the number of relevant parameters,
however, does not seem 
possible in the Euler-angle 
parameterization case. So, the analogue of the separability
function we will obtain here will be a {\it trivariate} function (of the
three eigenvalues). Hopefully, we will be able to determine an exact 
functional form for it, and then utilize it in further integrations
to obtain separable volumes, in terms of both 
monotone and non-monotone metrics. (Also, the question of whether counterpart 
Euler-angle separability
functions in the real and quaternionic cases adhere to 
some form of Dyson-index-sequence 
behavior certainly merits attention.)
\subsection{{\it Trivariate} 
separability function for {\it volume} computation}
Now, we made use of a sequence of 1,900,000 
12-dimensional Tezuka-Faure points (twelve being the number of 
Euler angles over which we will integrate).
For each such point, we let the associated three (free) eigenvalues 
each take on 
all possible values from $\frac{1}{40}$ to 1, in steps of $\frac{1}{40}$. Of 
course, the possible triads of free 
eigenvalues is constrained by the requirement
that they not all 
sum to more than 1. There were 9,880 such possible triads. 
For each such triple--holding the Euler angles constant--we 
evaluated whether the associated
$4 \times 4$ density matrix was separable or not.

\begin{figure}
\includegraphics{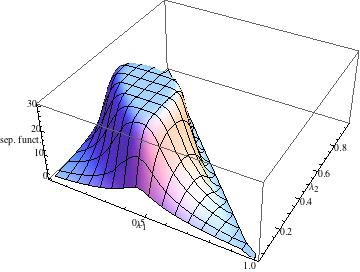}
\caption{\label{fig:TBS2}Two-dimensional marginal 
section of the estimated
three-dimensional separability function based on the Euler-angle
parameterization for the 15-dimensional convex set of complex 
$4 \times 4$ density matrices. Note the mesa/plateau shape, indicative of
the fully separable neighborhood of the fully-mixed state}
\end{figure}

We, then, interpolated the results to obtain functions defined over the
three-dimensional hypercube $[0,1]^3$.
In Fig.~\ref{fig:TBS2}, we show a two-dimensional marginal section 
(over $\lambda_1, \lambda_2$, say) of this function 
(obtained by summing over the values 
of $\lambda_3$) of the 
estimated three-dimensional separability function. 
We know from the work of Pittenger and Rubin \cite[Cor. 4.2]{pittenger}, for 
example,  that, 
for the specific case of two-qubits, any density matrix all of the 
four eigenvalues
of which are greater than $\frac{7}{30} \approx 0.2333$ {\it must} 
be separable.
Therefore, there certainly does exist 
a neighborhood of the fully-mixed state (having its
four eigenvalues equal to 
$\frac{1}{4}$) that is 
composed of only 
separable states. 
This is reflected in the plateau present in Fig..~\ref{fig:TBS2}.

\begin{figure}
\includegraphics{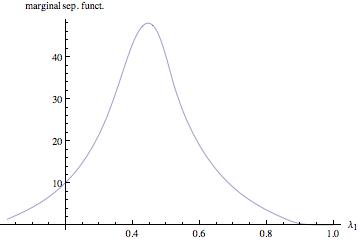}
\caption{\label{fig:TBS1}One-dimensional marginal section of the estimated
three-dimensional separability function based on the Euler-angle
parameterization for the 15-dimensional convex set of complex
$4 \times 4$ density matrices}
\end{figure}

In Fig.~\ref{fig:TBS1} we, additionally display a one-dimensional 
marginal section (over $\lambda_1$), obtained by summing 
over both
$\lambda_1$ and $\lambda_2$, of the 
estimated three-dimensional separability function. (The curve now appears
unimodal rather than flat at its maximum,)

Employing the trivariate separability function obtained by interpolation
from the data we have generated here, we were able to obtain an estimate of
0.242021 for the HS separability probability. (From our extensive 
Bloore-parameterization analyses, as previously noted 
we believe its true value is 
$\frac{8}{33} \approx 
0.242424$.)
We were essentially just 
as readily able to obtain an estimate of the Bures 
(minimal monotone) separability probability (or any of the other monotone 
metrics--Kubo-Mori, Wigner-Yanse,\ldots \cite{petzsudar}, 
it appears) of 0.0734223, while in
\cite{slaterJGP}, this had been conjectured to equal
$\frac{1680 \left(-1+\sqrt{2}\right)}{\pi ^8} \approx 0.733389$. 
(In the HS and Bures computations reported here and in the next section, 
we perform numerical integrations over the simplices of eigenvalues, 
in which the integrands are the products of our interpolated separability
functions and the appropriate 
scenario-specific volume elements indicated in the 
twin Sommers-\.Zyczkowski 2003 papers \cite{szHS,szBures}.)

Of course, now the research agenda should turn to the issue of finding a
possibly exact formula (undoubtedly {\it symmetric} in the three eigenvalues 
\cite[secs. V and VI]{slaterJPAreject}) 
for this three-dimensional Euler-angle-based
separability function, and for other qubit-qubit and 
qubit-qutrit scenarios.

\begin{figure}
\includegraphics{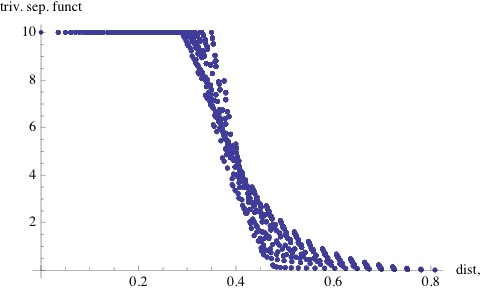}
\caption{\label{fig:SepFunctMassTri}The trivariate Euler-angle
complex qubit-qubit 
separability function paired with the 
{\it Euclidean distance} 
of the corresponding vector of eigenvalues 
to the vector ($\frac{1}{4},\frac{1}{4},\frac{1}{4},\frac{1}{4})$,
corresponding to the fully mixed state}
\end{figure}

In Fig.~\ref{fig:SepFunctMassTri} we show, based on the 9,880 points 
sampled for each 12-dimensional TF-point, 
the estimated value of the separability function for that point {\it 
paired} with
the Euclidean distance of the vector of eigenvalues 
($\lambda_1,\lambda_2,\lambda_3,1-\lambda_1-\lambda_2-\lambda_3$) 
for that point from
the vector of eigenvalues 
($\frac{1}{4},\frac{1}{4},\frac{1}{4},\frac{1}{4})$, 
corresponding to the fully mixed state.
\subsection{{\it  Bivariate} separability function for {\it area} computation} 
\label{secBivariate}
\begin{figure}
\includegraphics{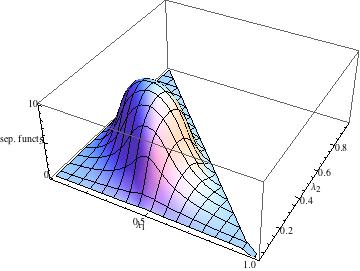}
\caption{\label{fig:TBSarea2}Bivariate Euler-angle 
separability function for the 14-dimensional boundary hyperarea
of the 15-dimensional convex set of complex
$4 \times 4$ density matrices}
\end{figure}

\begin{figure}
\includegraphics{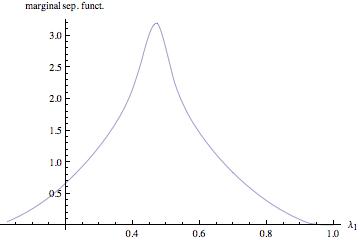}
\caption{\label{fig:TBSarea1}One-dimensional marginal section
of the estimated
two-dimensional Euler-angle-based separability function for the 
14-dimensional hyperarea of the 15-dimensional convex set of complex
$4 \times 4$ density matrices}
\end{figure}

Now we repeat the procedures described immediately before, except for
the {\it a priori} setting of {\it one} 
of the three free eigenvalues to zero, so the  
associated density matrices must lie on the 14-dimensional boundary
of the 15-dimensional convex set of two-qubit complex states.
(The analysis was conducted independently of that pertaining to the volume,
and now we were able to employ a much larger number--23,500,000--of TF-points.)
The resulting separability function is now bivariate, lending itself
immediately to graphic display.
In Fig.~\ref{fig:TBSarea2} we show this 
function, and in Fig.~\ref{fig:TBSarea1}, its one-dimensional section
over $\lambda_1$.

Employing the bivariate separability function (Fig.~\ref{fig:TBSarea2}) 
obtained by interpolation
from the data (23,500,000 TF-points) 
we generated, we were able to obtain an estimate of
0.12119 for the HS separability probability, which, from our
complementary Bloore analyses, together with the (``one-half'') 
Theorem 2 of 
\cite{sbz}, we believe to be exactly $\frac{4}{33} \approx
0.121212$. (The proximity of our estimate to this value clearly serves
to further fortify our conjecture that the HS separability probability of
generic complex two-qubit states is $\frac{8}{33}$.)
Additionally, our estimate of the associated {\it Bures} 
separability probability was 0.0396214, {\it approximately} one-half that of
the corresponding probability for the non-degenerate 
complex two-qubit states 
\cite{slaterPRA}.

\begin{figure}
\includegraphics{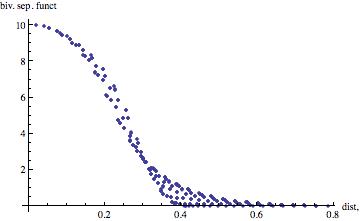}
\caption{\label{fig:SepFunctMassBi}Paired values of the estimated 
bivariate Euler-angle separability function (Fig.~\ref{fig:TBSarea2}) 
and the 
{\it Euclidean distance} of the associated
vector of eigenvalues to the vector ($\frac{1}{3},\frac{1}{3},\frac{1}{3},0)$
corresponding to the most fully mixed boundary state}
\end{figure}

In Fig.~\ref{fig:SepFunctMassBi} we show, based on 780 points sampled 
for each 12-dimensional TF-point, 
the estimated value of the bivariate separability function for that 
point paired with the Euclidean distance of the vector of eigenvalues 
for that 
point from the vector of eigenvalues 
($\frac{1}{3},\frac{1}{3},\frac{1}{3},0)$ 
corresponding to the most fully mixed boundary state. 
\subsection{Participation ratios}
\begin{figure}
\includegraphics{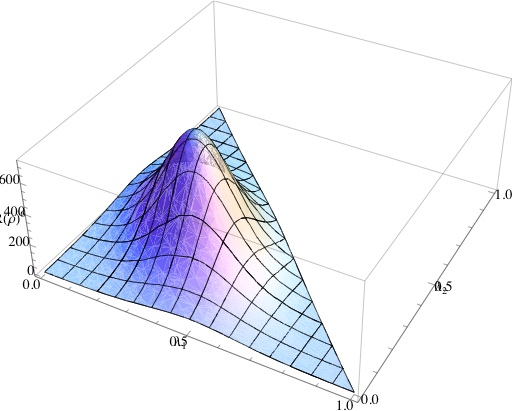}
\caption{\label{fig:participationRatio}The sixth-power, 
$R(\rho)^{6}$, of the participation ratio 
for minimally degenerate two-qubit states 
with one eigenvalue equal to zero. Note the similarity to Fig.~\ref{fig:TBSarea2}}
\end{figure}

\begin{figure}
\includegraphics{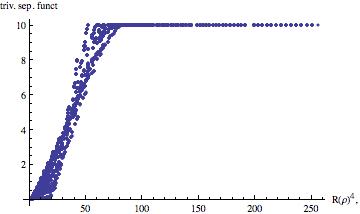}
\caption{\label{fig:partTri}The (Euler-angle) 
trivariate separability function 
for generic (15-dimensional) complex two-qubit states plotted
against the fourth-power 
of the participation ratio for the 9,880 points 
sampled}
\end{figure}

\begin{figure}
\includegraphics{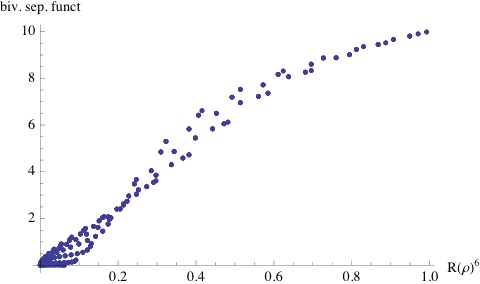}
\caption{\label{fig:partBi}The (Euler-angle)
bivariate separability function
for generic (14-dimensional) 
minimally degenerate complex two-qubit states plotted
against the sixth-power 
of the participation ratio for the 780 points
sampled}
\end{figure}

The {\it participation ratio} of a state $\rho$ is defined 
as  \cite[eq. (17)]{ZHSL} \cite[eq. (15.61)]{ingemarkarol} 
\cite{batlecasas1,batleplastino1}
\begin{equation}
R(\rho)= \frac{1}{\mbox{tr} \rho^2} .
\end{equation}
For $R(\rho)>3$, a two-qubit state {\it must} be separable. 
For convenience, we will also employ the variable
\begin{equation}
S(\rho)=\frac{3}{2} (1-\frac{1}{R(\rho)}),
\end{equation}
which varies over the interval [0,1] for states {\it outside} the separable 
ball.
In Fig.~\ref{fig:participationRatio} we show a plot--having set {\it one} 
of the four eigenvalues to zero--of 
the sixth-power 
of this ratio. We note a close similarity in its shape to the 
estimated bivariate
separability function displayed in
Fig.~\ref{fig:TBSarea2}.
Although we can not similarly visually display 
the trivariate separability function we have found
that it is closely fit--outside the separable ball ($R(\rho)>3$) 
\cite[Fig. 15.7]{ingemarkarol}-- by the fourth-power
of the participation ratio.
In Fig.~\ref{fig:partTri}, we show the trivariate separability function
{\it vs.} $R(\rho)^{4}$. (For $R(\rho)^{4}> 
81$, 
only separable states are encountered).
In Fig.~\ref{fig:partBi}, we show the comparable 
bivariate separability function
{\it vs.} $R(\rho)^{6}$.

As an exercise (not directly tied to our quasi-Monte Carlo computations), 
we assumed that the trivariate Euler-angle separability functions are  all 
proportional (outside the separable ball, $R(\rho) >3$) 
to some powers of $R(\rho)$, for each of the generic real, 
complex, truncated quaternionic and quaternionic two-qubit states.
Then, we found those powers which fit our conjectured values 
(discussed in sec.~\ref{Bp}) 
for the associated Hilbert-Schmidt two-qubit separability probabilities. 
The powers we found were 1.36743 ($\beta=1$), 2.36904 ($\beta=2$) 
and 4.0632 ($\beta=4$). We observe here a  
rough approximation to Dyson-index behavior, but can speculate that
when and if the true forms of the Euler-angle separability functions are found, such behavior will be strictly adhered to. (In fact, one research strategy 
might be to seek functions that fully conform to this principle, 
while fitting the conjectured HS separability probabilities. 
Also, below we will find closer adherence when we switch from the use
of the participation ratio to a simple linear transform of the 
Verstraete-Audenaert-De Moor-function \cite{ver} (cf. \cite{roland}), which
provides an improved bound on separability.)

When we similarly 
sought to fit our prediction of $\frac{4}{33}$ (that is, one-half of 
$\frac{8}{33}$ by the results of \cite{sbz}) for the HS
separability probability of generic minimally degenerate complex two-qubit
states to a bivariate function proportional to a power of
the participation ratio, we obtained a power of 6.11646, according rather 
well with Fig.~\ref{fig:partBi}.

In terms of the Hilbert-Schmidt metric, the lower bound on the 
complex two-qubit separability
probability provided by the separable ball ($R(\rho)>3$) 
is rather small, that is
$\frac{35 \pi}{23328 \sqrt{3}} \approx 0.00272132$, while relying upon 
the improved inequality  reported in \cite{ver},
\begin{equation} \label{VADbound}
VAD(\rho)= 
\lambda_{1}-\lambda_3 -2 \sqrt{\lambda_2 \lambda_4} <0,\hspace{.3in}
(\lambda_1>\lambda_2 >\lambda_3 >\lambda_4),
\end{equation}
 it is 0.00365406. These figures are both considerably smaller
than the comparable ones (0.3023 and 0.3270) 
given in \cite{ver} using (apparently) 
the measure ({\it uniform} on the simplex of eigenvalues) first employed in
\cite{ZHSL}.
\subsection{Verstraete-Audenaert-De Moor function}
If we switch from the participation ratio 
$R(\rho)$ to a simple linear transformation of the 
 Verstraete-Audenaert-De Moor function, that is, 
$1-VAD(\rho)$ (which varies over [0,1] for states outside the separable 
VAD set), in seeking to fit the trivariate
separability function to our conjectured Hilbert-Schmidt two-qubit 
separability 
probabilities (sec.~\ref{Bp}), we find that 
in the generic complex ($\beta=2$) case, 
$\Big(1-VAD(\rho)\Big)^{3.15448}$ gives the best fit (for $VAD(\rho)>0$).
Then, closely consistent with Dyson-index behavior, we obtained
$\Big(1-VAD(\rho)\Big)^{1.53785}$ as the best fit in the generic real ($\beta=1$) 
scenario. (The VAD-bound (\ref{VADbound}) 
provides us with no useful information if we set $\lambda_4=0$, so no 
relevance to the minimally degenerate two-qubit scenario is apparent.)

\subsection{Beta function fits to Euler-angle separability functions}
\begin{figure}
\includegraphics{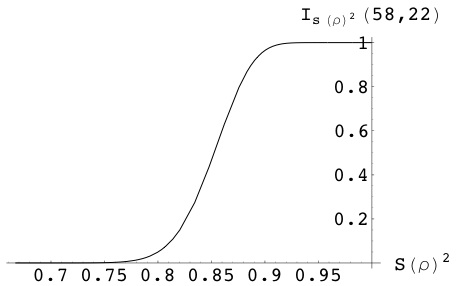}
\caption{\label{fig:BetaFit1}Incomplete beta function estimate of the
{\it bivariate} Euler-angle separability function that closely 
reproduces the  conjectured Hilbert-Schmidt
{\it minimally degenerate} 
two-qubit real, complex and quaternionic
separability probabilities}
\end{figure}

We can fit within $0.4\%$ 
our conjectured values of $\frac{4}{33}$ and $\frac{4}{17}$ for
the Hilbert-Schmidt separability probabilities of the complex and real
minimally degenerate two-qubit states, respectively, by assuming--in line with the Dyson-index ansatz-- that 
the Euler-angle separability function in the {\it real} case is a 
{\it regularized beta function} (incomplete beta function ratio) 
\cite[p. 11]{handbook} of the form
$I_{S(\rho)^2}(58,22)$ (Fig.~\ref{fig:BetaFit1}), 
{\it and} the Euler-angle separability function 
(Fig.~\ref{fig:TBSarea2}) 
in the {\it complex} case, the {\it square} of that function. 
(Continuing along such lines, if we employ the fourth-power of the function,
our estimate of the associated quaternionic separability probability is some
$91.45\%$ of the conjectured value of $\frac{36221472}{936239725} 
\approx 0.0386882$.)

\begin{figure}
\includegraphics{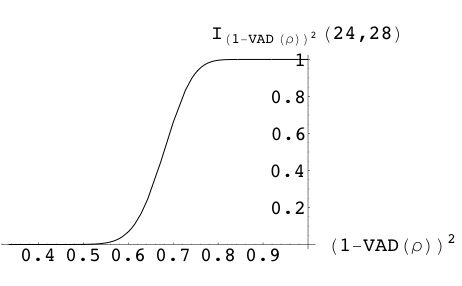}
\caption{\label{fig:BetaFit2}Incomplete beta function estimate of the
{\it trivariate} Euler-angle separability function that closely
reproduces the  conjectured Hilbert-Schmidt
{\it nondegenerate} 
two-qubit real, complex and quaternionic separability probabilities}
\end{figure}

Similarly, for the generic {\it nondegenerate} complex and real
two-qubit states, we can achieve fits within $0.7\%$ to {\it both} 
the conjectured
HS separability probabilities of $\frac{8}{33}$ and $\frac{8}{17}$, 
respectively, by taking in the real case the Euler-angle separability
function to be $I_{\Big((1-VAD(\rho)\Big)^2}(24,28)$ 
(Fig.~\ref{fig:BetaFit2}) and its square in the 
complex case. (Use of its fourth power to estimate the
HS {\it quaternionic} two-qubit separability probability yielded a result 
0.795969 as large as the value, 
$\frac{72442944}{936239725}
\approx 0.0795969$, conjectured above (\ref{HSquat}).)

\section{Summary}
We have extended the findings and analyses of our two recent studies 
\cite{slaterPRA2} and \cite{slater833} by, first, 
obtaining numerical estimates of the separability function based on the 
(Euclidean, flat) Hilbert-Schmidt (HS) 
metric for the 27-dimensional convex set of 
quaternionic two-qubit systems (sec.~\ref{secQuat}). 
The estimated function closely conformed to
our previously-formulated Dyson-index ($\beta = 1, 2, 4$) 
ansatz, dictating that 
the quaternionic ($\beta=4$) separability function should be
exactly proportional to the square of the separability function
for the 
15-dimensional convex set of 
two-qubit complex ($\beta=2$) 
systems, as well as the fourth power of the 
separability function for the 9-dimensional convex set of 
two-qubit real ($\beta=1$) systems. 
In particular, these additional analyses led us specifically to aver that 
\begin{equation}
\mathcal{S}^{HS}_{quat}(\mu) = (\frac{6}{71})^2 \Big((3-\mu^2) \mu\Big)^4
= (\mathcal{S}^{HS}_{complex}(\mu))^2, 0 \leq \mu \leq 1.
\end{equation} Here, 
$\mu =\sqrt{\frac{\rho_{11} \rho_{44}}{\rho_{22}
\rho_{33}}}$,
where $\rho$ denotes a $4 \times 4$ two-qubit density matrix. 
We have, thus, been able to supplement ({\it and} fortify) 
our previous assertion that the HS separability
probability of the two-qubit complex
states is $\frac{8}{33} \approx 0.242424$, claiming that its quaternionic 
counterpart is $\frac{72442944}{936239725} \approx 0.0773765$. 
We have also commented on and analyzed the odd $\beta=1$ and $\beta=3$ cases 
(sec.~\ref{secTrunc}), which still remain somewhat problematical.

Further, we found (sec.~\ref{QubQut}) 
strong evidence of adherence to the Dyson-index 
ansatz for the 25-dimensional real and 35-dimensional complex
qubit-{\it qutrit} systems
with real HS separability function being proportional to a function of 
the form,
$1-(1-\nu_1 \nu_2)^{{\frac{5}{2}}}$, with $\nu_1,\nu_2$ defined in 
(\ref{tworatios}). 
Subject to the validity of this separability function, 
we have obtained the corresponding $R_{2}$ constants
($\beta=1,\ldots,4)$ and estimated the complementary $R_{1}$ constants
(the products $R_{1} R_{2}$ giving throughout, in our 
fundamental paradigm,
the desired separability probabilities).

Then (sec.~\ref{secBures}), 
we determined that in terms of the
Bures (minimal monotone) metric--for certain, basic simple scenarios 
(in which the diagonal entries of $\rho$ are
unrestricted, and one or two off-diagonal
[real, complex or quaternionic] pairs of entries are nonzero)--that the
Dyson-index power relations no longer strictly hold, but
come remarkably close to doing so.

Finally (sec.~\ref{secEuler}), we examined the possibility of defining
``separability functions'' using the Euler-angle parameterization
of Tilma, Byrd and Sudarshan \cite{tbs}, 
rather than the Bloore (correlation/off-diagonal scaling) 
framework \cite{bloore}. Although now we are, {\it prima 
facie}, faced with a trivariate (in the complex two-qubit case) 
function, rather than a univariate one, 
we do not encounter the problem of having to 
determine an overall normalization
factor for the separability function, since it is known that any density
matrix with all eigenvalues equal to one another must be separable.
It also appears that this simplifying feature further extends to the case
where minimally degenerate (boundary) complex qubit-qubit 
states are considered (sec.~\ref{secBivariate}), with any
such state having its three non-zero eigenvalues all equal to $\frac{1}{3}$ 
(lying on the boundary of the separable ball $R(\rho)=3$)
being necessarily separable.
Use of the estimated 
Euler-angle trivariate separability function lent still further 
(numerical) support
to the $\frac{8}{33}$ conjecture \cite{slater833} 
for the HS separability probability (and the associated $\frac{4}{33}$ 
conjecture for the hyperarea of the minimally degenerate two-qubit states)
and the $\frac{1680 (\sqrt{2}-1)}{\pi^8}$ ``silver mean'' conjecture 
\cite{slaterJGP} for 
the Bures separability probability 
of generic complex two-qubit states.
\begin{acknowledgments}
I would like to express appreciation to the Kavli Institute for Theoretical
Physics (KITP)
for computational support in this research. Also K, 
{\,Z}yczkowski alerted me 
to the relevance of \cite{ver}, and M. Trott assisted with certain
computations.

\end{acknowledgments}

\bibliography{EulerSep6}% Produces the bibliography via BibTeX.

\begin{thebibliography}{70}
\expandafter\ifx\csname natexlab\endcsname\relax\def\natexlab#1{#1}\fi
\expandafter\ifx\csname bibnamefont\endcsname\relax
  \def\bibnamefont#1{#1}\fi
\expandafter\ifx\csname bibfnamefont\endcsname\relax
  \def\bibfnamefont#1{#1}\fi
\expandafter\ifx\csname citenamefont\endcsname\relax
  \def\citenamefont#1{#1}\fi
\expandafter\ifx\csname url\endcsname\relax
  \def\url#1{\texttt{#1}}\fi
\expandafter\ifx\csname urlprefix\endcsname\relax\def\urlprefix{URL }\fi
\providecommand{\bibinfo}[2]{#2}
\providecommand{\eprint}[2][]{\url{#2}}

\bibitem[{\citenamefont{{\.Z}yczkowski
  et~al.}(1998)\citenamefont{{\.Z}yczkowski, Horodecki, Sanpera, and
  Lewenstein}}]{ZHSL}
\bibinfo{author}{\bibfnamefont{K.}~\bibnamefont{{\.Z}yczkowski}},
  \bibinfo{author}{\bibfnamefont{P.}~\bibnamefont{Horodecki}},
  \bibinfo{author}{\bibfnamefont{A.}~\bibnamefont{Sanpera}}, \bibnamefont{and}
  \bibinfo{author}{\bibfnamefont{M.}~\bibnamefont{Lewenstein}},
  \bibinfo{journal}{Phys. Rev. A} \textbf{\bibinfo{volume}{58}},
  \bibinfo{pages}{883} (\bibinfo{year}{1998}).

\bibitem[{\citenamefont{Avron et~al.}(2007)\citenamefont{Avron, Bisker, and
  Kenneth}}]{avron}
\bibinfo{author}{\bibfnamefont{J.~E.} \bibnamefont{Avron}},
  \bibinfo{author}{\bibfnamefont{G.}~\bibnamefont{Bisker}}, \bibnamefont{and}
  \bibinfo{author}{\bibfnamefont{O.}~\bibnamefont{Kenneth}},
  \bibinfo{journal}{J. Math. Phys.} \textbf{\bibinfo{volume}{48}},
  \bibinfo{pages}{102107} (\bibinfo{year}{2007}).

\bibitem[{\citenamefont{Tilma and Sudarshan}(2002)}]{sudarshan}
\bibinfo{author}{\bibfnamefont{T.}~\bibnamefont{Tilma}} \bibnamefont{and}
  \bibinfo{author}{\bibfnamefont{E.~C.~G.} \bibnamefont{Sudarshan}},
  \bibinfo{journal}{J. Phys. A} \textbf{\bibinfo{volume}{35}},
  \bibinfo{pages}{10467} (\bibinfo{year}{2002}).

\bibitem[{\citenamefont{Slater}(1999{\natexlab{a}})}]{slaterHall}
\bibinfo{author}{\bibfnamefont{P.~B.} \bibnamefont{Slater}},
  \bibinfo{journal}{J. Phys. A} \textbf{\bibinfo{volume}{32}},
  \bibinfo{pages}{8231} (\bibinfo{year}{1999}{\natexlab{a}}).

\bibitem[{\citenamefont{Slater}(1999{\natexlab{b}})}]{slaterA}
\bibinfo{author}{\bibfnamefont{P.~B.} \bibnamefont{Slater}},
  \bibinfo{journal}{J. Phys. A} \textbf{\bibinfo{volume}{32}},
  \bibinfo{pages}{5261} (\bibinfo{year}{1999}{\natexlab{b}}).

\bibitem[{\citenamefont{Slater}(2000{\natexlab{a}})}]{slaterC}
\bibinfo{author}{\bibfnamefont{P.~B.} \bibnamefont{Slater}},
  \bibinfo{journal}{Euro. Phys. J. B} \textbf{\bibinfo{volume}{17}},
  \bibinfo{pages}{471} (\bibinfo{year}{2000}{\natexlab{a}}).

\bibitem[{\citenamefont{Slater}(2000{\natexlab{b}})}]{slaterOptics}
\bibinfo{author}{\bibfnamefont{P.~B.} \bibnamefont{Slater}},
  \bibinfo{journal}{J. Opt. B} \textbf{\bibinfo{volume}{2}},
  \bibinfo{pages}{L19} (\bibinfo{year}{2000}{\natexlab{b}}).

\bibitem[{\citenamefont{Slater}(2005{\natexlab{a}})}]{slaterJGP}
\bibinfo{author}{\bibfnamefont{P.~B.} \bibnamefont{Slater}},
  \bibinfo{journal}{J. Geom. Phys.} \textbf{\bibinfo{volume}{53}},
  \bibinfo{pages}{74} (\bibinfo{year}{2005}{\natexlab{a}}).

\bibitem[{\citenamefont{Slater}(2005{\natexlab{b}})}]{slaterPRA}
\bibinfo{author}{\bibfnamefont{P.~B.} \bibnamefont{Slater}},
  \bibinfo{journal}{Phys. Rev. A} \textbf{\bibinfo{volume}{71}},
  \bibinfo{pages}{052319} (\bibinfo{year}{2005}{\natexlab{b}}).

\bibitem[{\citenamefont{Slater}(2006)}]{pbsCanosa}
\bibinfo{author}{\bibfnamefont{P.~B.} \bibnamefont{Slater}},
  \bibinfo{journal}{J. Phys. A} \textbf{\bibinfo{volume}{39}},
  \bibinfo{pages}{913} (\bibinfo{year}{2006}).

\bibitem[{\citenamefont{Slater}(2007{\natexlab{a}})}]{slaterPRA2}
\bibinfo{author}{\bibfnamefont{P.~B.} \bibnamefont{Slater}},
  \bibinfo{journal}{Phys. Rev. A} \textbf{\bibinfo{volume}{75}},
  \bibinfo{pages}{032326} (\bibinfo{year}{2007}{\natexlab{a}}).

\bibitem[{\citenamefont{Kass}(1989)}]{kass}
\bibinfo{author}{\bibfnamefont{R.~E.} \bibnamefont{Kass}},
  \bibinfo{journal}{Statist. Sci.} \textbf{\bibinfo{volume}{4}},
  \bibinfo{pages}{188} (\bibinfo{year}{1989}).

\bibitem[{\citenamefont{{\.Z}yczkowski and Sommers}(2003)}]{szHS}
\bibinfo{author}{\bibfnamefont{K.}~\bibnamefont{{\.Z}yczkowski}}
  \bibnamefont{and} \bibinfo{author}{\bibfnamefont{H.-J.}
  \bibnamefont{Sommers}}, \bibinfo{journal}{J. Phys. A}
  \textbf{\bibinfo{volume}{36}}, \bibinfo{pages}{10115} (\bibinfo{year}{2003}).

\bibitem[{\citenamefont{Sommers and {\.Z}yczkowski}(2003)}]{szBures}
\bibinfo{author}{\bibfnamefont{H.-J.} \bibnamefont{Sommers}} \bibnamefont{and}
  \bibinfo{author}{\bibfnamefont{K.}~\bibnamefont{{\.Z}yczkowski}},
  \bibinfo{journal}{J. Phys. A} \textbf{\bibinfo{volume}{36}},
  \bibinfo{pages}{10083} (\bibinfo{year}{2003}).

\bibitem[{\citenamefont{Mehta}(2004)}]{random}
\bibinfo{author}{\bibfnamefont{M.~L.} \bibnamefont{Mehta}},
  \emph{\bibinfo{title}{Random Matrices}}
  (\bibinfo{publisher}{Elsevier/Academic}, \bibinfo{address}{Amsterdam},
  \bibinfo{year}{2004}).

\bibitem[{\citenamefont{Andai}(2006)}]{andai}
\bibinfo{author}{\bibfnamefont{A.}~\bibnamefont{Andai}}, \bibinfo{journal}{J.
  Phys. A} \textbf{\bibinfo{volume}{39}}, \bibinfo{pages}{13641}
  (\bibinfo{year}{2006}).

\bibitem[{\citenamefont{Slater}(2007{\natexlab{b}})}]{slater833}
\bibinfo{author}{\bibfnamefont{P.~B.} \bibnamefont{Slater}},
  \bibinfo{journal}{J. Phys. A} \textbf{\bibinfo{volume}{40}},
  \bibinfo{pages}{14279} (\bibinfo{year}{2007}{\natexlab{b}}).

\bibitem[{\citenamefont{Slater}({\natexlab{a}})}]{slaterDyson}
\bibinfo{author}{\bibfnamefont{P.~B.} \bibnamefont{Slater}},
  \eprint{arXiv:0708,4208}.

\bibitem[{\citenamefont{Bloore}(1976)}]{bloore}
\bibinfo{author}{\bibfnamefont{F.~J.} \bibnamefont{Bloore}},
  \bibinfo{journal}{J. Phys. A} \textbf{\bibinfo{volume}{9}},
  \bibinfo{pages}{2059} (\bibinfo{year}{1976}).

\bibitem[{\citenamefont{Joe}(2006)}]{joe}
\bibinfo{author}{\bibfnamefont{H.}~\bibnamefont{Joe}}, \bibinfo{journal}{J.
  Multiv. Anal.} \textbf{\bibinfo{volume}{97}}, \bibinfo{pages}{2177}
  (\bibinfo{year}{2006}).

\bibitem[{\citenamefont{Kurowicka and Cooke}(2003)}]{kurowicka}
\bibinfo{author}{\bibfnamefont{D.}~\bibnamefont{Kurowicka}} \bibnamefont{and}
  \bibinfo{author}{\bibfnamefont{R.}~\bibnamefont{Cooke}},
  \bibinfo{journal}{Lin. Alg. Applics.} \textbf{\bibinfo{volume}{372}},
  \bibinfo{pages}{225} (\bibinfo{year}{2003}).

\bibitem[{\citenamefont{Kurowicka and Cooke}(2006)}]{kurowicka2}
\bibinfo{author}{\bibfnamefont{D.}~\bibnamefont{Kurowicka}} \bibnamefont{and}
  \bibinfo{author}{\bibfnamefont{R.~M.} \bibnamefont{Cooke}},
  \bibinfo{journal}{Lin. Alg. Applics.} \textbf{\bibinfo{volume}{418}},
  \bibinfo{pages}{188} (\bibinfo{year}{2006}).

\bibitem[{\citenamefont{Guiasu}(1987)}]{guiasu}
\bibinfo{author}{\bibfnamefont{S.}~\bibnamefont{Guiasu}},
  \bibinfo{journal}{Phys. Rev. A} \textbf{\bibinfo{volume}{36}},
  \bibinfo{pages}{1971} (\bibinfo{year}{1987}).

\bibitem[{\citenamefont{G{\"u}hne et~al.}(2007)\citenamefont{G{\"u}hne, Hyllus,
  Gittsovich, and Eisert}}]{guhne}
\bibinfo{author}{\bibfnamefont{O.}~\bibnamefont{G{\"u}hne}},
  \bibinfo{author}{\bibfnamefont{P.}~\bibnamefont{Hyllus}},
  \bibinfo{author}{\bibfnamefont{O.}~\bibnamefont{Gittsovich}},
  \bibnamefont{and} \bibinfo{author}{\bibfnamefont{J.}~\bibnamefont{Eisert}},
  \bibinfo{journal}{Phys. Rev. Lett.} \textbf{\bibinfo{volume}{99}},
  \bibinfo{pages}{130504} (\bibinfo{year}{2007}).

\bibitem[{\citenamefont{Mkrtchian and Chaltykyan}(1987)}]{vanik}
\bibinfo{author}{\bibfnamefont{V.~E.} \bibnamefont{Mkrtchian}}
  \bibnamefont{and} \bibinfo{author}{\bibfnamefont{V.~O.}
  \bibnamefont{Chaltykyan}}, \bibinfo{journal}{Opt. Commun.}
  \textbf{\bibinfo{volume}{63}}, \bibinfo{pages}{239} (\bibinfo{year}{1987}).

\bibitem[{\citenamefont{Peres}(1979)}]{asher2}
\bibinfo{author}{\bibfnamefont{A.}~\bibnamefont{Peres}},
  \bibinfo{journal}{Phys. Rev. Lett.} \textbf{\bibinfo{volume}{42}},
  \bibinfo{pages}{683} (\bibinfo{year}{1979}).

\bibitem[{\citenamefont{Adler}(1995)}]{adler}
\bibinfo{author}{\bibfnamefont{S.~L.} \bibnamefont{Adler}},
  \emph{\bibinfo{title}{Quaternionic quantum mechanics and quantum fields}}
  (\bibinfo{publisher}{Oxford}, \bibinfo{address}{New York},
  \bibinfo{year}{1995}).

\bibitem[{\citenamefont{Batle et~al.}(2003)\citenamefont{Batle, Plastino,
  Casas, and Plastino}}]{batle2}
\bibinfo{author}{\bibfnamefont{J.}~\bibnamefont{Batle}},
  \bibinfo{author}{\bibfnamefont{A.~R.} \bibnamefont{Plastino}},
  \bibinfo{author}{\bibfnamefont{M.}~\bibnamefont{Casas}}, \bibnamefont{and}
  \bibinfo{author}{\bibfnamefont{A.}~\bibnamefont{Plastino}},
  \bibinfo{journal}{Opt. Spect.} \textbf{\bibinfo{volume}{94}},
  \bibinfo{pages}{1562} (\bibinfo{year}{2003}).

\bibitem[{\citenamefont{Peres}(1996)}]{asher}
\bibinfo{author}{\bibfnamefont{A.}~\bibnamefont{Peres}},
  \bibinfo{journal}{Phys. Rev. Lett.} \textbf{\bibinfo{volume}{77}},
  \bibinfo{pages}{1413} (\bibinfo{year}{1996}).

\bibitem[{\citenamefont{Horodecki et~al.}(1996)\citenamefont{Horodecki,
  Horodecki, and Horodecki}}]{michal}
\bibinfo{author}{\bibfnamefont{M.}~\bibnamefont{Horodecki}},
  \bibinfo{author}{\bibfnamefont{P.}~\bibnamefont{Horodecki}},
  \bibnamefont{and}
  \bibinfo{author}{\bibfnamefont{R.}~\bibnamefont{Horodecki}},
  \bibinfo{journal}{Phys. Lett. A} \textbf{\bibinfo{volume}{223}},
  \bibinfo{pages}{1} (\bibinfo{year}{1996}).

\bibitem[{\citenamefont{Tilma et~al.}(2002)\citenamefont{Tilma, Byrd, and
  Sudarshan}}]{tbs}
\bibinfo{author}{\bibfnamefont{T.}~\bibnamefont{Tilma}},
  \bibinfo{author}{\bibfnamefont{M.}~\bibnamefont{Byrd}}, \bibnamefont{and}
  \bibinfo{author}{\bibfnamefont{E.~C.~G.} \bibnamefont{Sudarshan}},
  \bibinfo{journal}{J. Phys. A} \textbf{\bibinfo{volume}{35}},
  \bibinfo{pages}{10445} (\bibinfo{year}{2002}).

\bibitem[{\citenamefont{Gupta and Nadarajah}(2004)}]{handbook}
\bibinfo{author}{\bibfnamefont{A.~K.} \bibnamefont{Gupta}} \bibnamefont{and}
  \bibinfo{author}{\bibfnamefont{S.}~\bibnamefont{Nadarajah}},
  \emph{\bibinfo{title}{Handbook of Beta Distribution and Its Applications}}
  (\bibinfo{publisher}{Marcel Dekker}, \bibinfo{address}{New York},
  \bibinfo{year}{2004}).

\bibitem[{\citenamefont{Dyson}(1970)}]{dyson}
\bibinfo{author}{\bibfnamefont{F.~J.} \bibnamefont{Dyson}},
  \bibinfo{journal}{Commun. Math. Phys.} \textbf{\bibinfo{volume}{19}},
  \bibinfo{pages}{235} (\bibinfo{year}{1970}).

\bibitem[{\citenamefont{Slater}({\natexlab{b}})}]{slaterJPAreject}
\bibinfo{author}{\bibfnamefont{P.~B.} \bibnamefont{Slater}},
  \eprint{quant-ph/0602109}.

\bibitem[{\citenamefont{Bellman}(1957)}]{bellman}
\bibinfo{author}{\bibfnamefont{R.}~\bibnamefont{Bellman}},
  \emph{\bibinfo{title}{Dynamic Programming}} (\bibinfo{publisher}{Princeton
  Univ.}, \bibinfo{address}{Princeton}, \bibinfo{year}{1957}).

\bibitem[{\citenamefont{Kuo and Sloan}(2005)}]{kuosloan}
\bibinfo{author}{\bibfnamefont{F.~Y.} \bibnamefont{Kuo}} \bibnamefont{and}
  \bibinfo{author}{\bibfnamefont{I.~H.} \bibnamefont{Sloan}},
  \bibinfo{journal}{Not. Amer. Math. Soc.} \textbf{\bibinfo{volume}{52}},
  \bibinfo{pages}{1320} (\bibinfo{year}{2005}).

\bibitem[{\citenamefont{{\"O}kten}(1999)}]{giray1}
\bibinfo{author}{\bibfnamefont{G.}~\bibnamefont{{\"O}kten}},
  \bibinfo{journal}{MATHEMATICA in Educ. Res.} \textbf{\bibinfo{volume}{8}},
  \bibinfo{pages}{52} (\bibinfo{year}{1999}).

\bibitem[{\citenamefont{Faure and Tezuka}(2002)}]{tezuka}
\bibinfo{author}{\bibfnamefont{H.}~\bibnamefont{Faure}} \bibnamefont{and}
  \bibinfo{author}{\bibfnamefont{S.}~\bibnamefont{Tezuka}}, in
  \emph{\bibinfo{booktitle}{Monte Carlo and Quasi-Monte Carlo Methods 2000
  (Hong Kong)}}, edited by \bibinfo{editor}{\bibfnamefont{K.~T.}
  \bibnamefont{Tang}}, \bibinfo{editor}{\bibfnamefont{F.~J.}
  \bibnamefont{Hickernell}}, \bibnamefont{and}
  \bibinfo{editor}{\bibfnamefont{H.}~\bibnamefont{Niederreiter}}
  (\bibinfo{publisher}{Springer}, \bibinfo{address}{Berlin},
  \bibinfo{year}{2002}), p. \bibinfo{pages}{242}.

\bibitem[{\citenamefont{Lov{\'a}sz and Vempala}(2006)}]{lovasz}
\bibinfo{author}{\bibfnamefont{L.}~\bibnamefont{Lov{\'a}sz}} \bibnamefont{and}
  \bibinfo{author}{\bibfnamefont{S.}~\bibnamefont{Vempala}},
  \bibinfo{journal}{J. Comput. Syst. Sci} \textbf{\bibinfo{volume}{72}},
  \bibinfo{pages}{392} (\bibinfo{year}{2006}).

\bibitem[{\citenamefont{Dyer et~al.}(1991)\citenamefont{Dyer, Frieze, and
  Kannan}}]{dyer}
\bibinfo{author}{\bibfnamefont{M.}~\bibnamefont{Dyer}},
  \bibinfo{author}{\bibfnamefont{A.}~\bibnamefont{Frieze}}, \bibnamefont{and}
  \bibinfo{author}{\bibfnamefont{R.}~\bibnamefont{Kannan}},
  \bibinfo{journal}{J. ACM} \textbf{\bibinfo{volume}{38}}, \bibinfo{pages}{1}
  (\bibinfo{year}{1991}).

\bibitem[{\citenamefont{Slater}(1996)}]{slaterJMP1996}
\bibinfo{author}{\bibfnamefont{P.~B.} \bibnamefont{Slater}},
  \bibinfo{journal}{J. Math. Phys.} \textbf{\bibinfo{volume}{37}},
  \bibinfo{pages}{2682} (\bibinfo{year}{1996}).

\bibitem[{\citenamefont{Jiang}(2005)}]{JIANG}
\bibinfo{author}{\bibfnamefont{T.}~\bibnamefont{Jiang}}, \bibinfo{journal}{J.
  Math. Phys} \textbf{\bibinfo{volume}{46}}, \bibinfo{pages}{052106}
  (\bibinfo{year}{2005}).

\bibitem[{\citenamefont{S{\'a}nchez-Ruiz}(1995)}]{sanchez}
\bibinfo{author}{\bibfnamefont{J.}~\bibnamefont{S{\'a}nchez-Ruiz}},
  \bibinfo{journal}{Phys. Rev. E} \textbf{\bibinfo{volume}{52}},
  \bibinfo{pages}{5653} (\bibinfo{year}{1995}).

\bibitem[{\citenamefont{Sen}(1996)}]{sen}
\bibinfo{author}{\bibfnamefont{S.}~\bibnamefont{Sen}}, \bibinfo{journal}{Phys.
  Rev. Lett.} \textbf{\bibinfo{volume}{77}}, \bibinfo{pages}{1}
  (\bibinfo{year}{1996}).

\bibitem[{\citenamefont{Pfaff}(2000)}]{pfaff}
\bibinfo{author}{\bibfnamefont{F.~R.} \bibnamefont{Pfaff}},
  \bibinfo{journal}{Amer. Math. Mon.} \textbf{\bibinfo{volume}{107}},
  \bibinfo{pages}{156} (\bibinfo{year}{2000}).

\bibitem[{\citenamefont{Budden et~al.}(2007)\citenamefont{Budden, Hadavas,
  Hoffman, and Pretz}}]{budden}
\bibinfo{author}{\bibfnamefont{M.}~\bibnamefont{Budden}},
  \bibinfo{author}{\bibfnamefont{P.}~\bibnamefont{Hadavas}},
  \bibinfo{author}{\bibfnamefont{L.}~\bibnamefont{Hoffman}}, \bibnamefont{and}
  \bibinfo{author}{\bibfnamefont{C.}~\bibnamefont{Pretz}},
  \bibinfo{journal}{Appl. Math. E-Notes} \textbf{\bibinfo{volume}{7}},
  \bibinfo{pages}{53} (\bibinfo{year}{2007}).

\bibitem[{\citenamefont{Bengtsson and {\.Z}yczkowski}(2006)}]{ingemarkarol}
\bibinfo{author}{\bibfnamefont{I.}~\bibnamefont{Bengtsson}} \bibnamefont{and}
  \bibinfo{author}{\bibfnamefont{K.}~\bibnamefont{{\.Z}yczkowski}},
  \emph{\bibinfo{title}{Geometry of Quantum States}}
  (\bibinfo{publisher}{Cambridge}, \bibinfo{address}{Cambridge},
  \bibinfo{year}{2006}).

\bibitem[{\citenamefont{Hall}(1998)}]{hall}
\bibinfo{author}{\bibfnamefont{M.~J.~W.} \bibnamefont{Hall}},
  \bibinfo{journal}{Phys. Lett. A} \textbf{\bibinfo{volume}{242}},
  \bibinfo{pages}{123} (\bibinfo{year}{1998}).

\bibitem[{\citenamefont{Slater}(1998)}]{slatersrednicki}
\bibinfo{author}{\bibfnamefont{P.~B.} \bibnamefont{Slater}},
  \bibinfo{journal}{Phys. Lett. A.} \textbf{\bibinfo{volume}{247}},
  \bibinfo{pages}{1} (\bibinfo{year}{1998}).

\bibitem[{\citenamefont{Asorey et~al.}(2007)\citenamefont{Asorey, Scolarici,
  and Solombrino}}]{asorey}
\bibinfo{author}{\bibfnamefont{M.}~\bibnamefont{Asorey}},
  \bibinfo{author}{\bibfnamefont{G.}~\bibnamefont{Scolarici}},
  \bibnamefont{and}
  \bibinfo{author}{\bibfnamefont{L.}~\bibnamefont{Solombrino}},
  \bibinfo{journal}{Phys. Rev. A} \textbf{\bibinfo{volume}{76}},
  \bibinfo{pages}{012111} (\bibinfo{year}{2007}).

\bibitem[{\citenamefont{Kogut et~al.}(2000)\citenamefont{Kogut, Stephanov,
  D.Toublan, Verbaarschot, and Zhitnitsky}}]{kogut}
\bibinfo{author}{\bibfnamefont{J.~B.} \bibnamefont{Kogut}},
  \bibinfo{author}{\bibfnamefont{M.~A.} \bibnamefont{Stephanov}},
  \bibinfo{author}{\bibnamefont{D.Toublan}},
  \bibinfo{author}{\bibfnamefont{J.~J.~M.} \bibnamefont{Verbaarschot}},
  \bibnamefont{and}
  \bibinfo{author}{\bibfnamefont{A.}~\bibnamefont{Zhitnitsky}},
  \bibinfo{journal}{Nucl. Phys. B} \textbf{\bibinfo{volume}{582}},
  \bibinfo{pages}{477} (\bibinfo{year}{2000}).

\bibitem[{\citenamefont{Caselle and Magnea}(2004)}]{caselle}
\bibinfo{author}{\bibfnamefont{M.}~\bibnamefont{Caselle}} \bibnamefont{and}
  \bibinfo{author}{\bibfnamefont{U.}~\bibnamefont{Magnea}},
  \bibinfo{journal}{Phys. Rep.} \textbf{\bibinfo{volume}{394}},
  \bibinfo{pages}{41} (\bibinfo{year}{2004}).

\bibitem[{\citenamefont{Dittmann}(1999)}]{explicit}
\bibinfo{author}{\bibfnamefont{J.}~\bibnamefont{Dittmann}},
  \bibinfo{journal}{J. Phys. A} \textbf{\bibinfo{volume}{32}},
  \bibinfo{pages}{2663} (\bibinfo{year}{1999}).

\bibitem[{\citenamefont{H{\"u}bner}(1992)}]{hubner}
\bibinfo{author}{\bibfnamefont{M.}~\bibnamefont{H{\"u}bner}},
  \bibinfo{journal}{Phys. Lett. A} \textbf{\bibinfo{volume}{163}},
  \bibinfo{pages}{239} (\bibinfo{year}{1992}).

\bibitem[{\citenamefont{Dittmann}(1993)}]{dittmann}
\bibinfo{author}{\bibfnamefont{J.}~\bibnamefont{Dittmann}},
  \bibinfo{journal}{Sem. Sophus Lie} \textbf{\bibinfo{volume}{3}},
  \bibinfo{pages}{73} (\bibinfo{year}{1993}).

\bibitem[{\citenamefont{Bru{\ss} and Macchiavello}(2005)}]{bruss}
\bibinfo{author}{\bibfnamefont{D.}~\bibnamefont{Bru{\ss}}} \bibnamefont{and}
  \bibinfo{author}{\bibfnamefont{C.}~\bibnamefont{Macchiavello}},
  \bibinfo{journal}{Found. Phys.} \textbf{\bibinfo{volume}{35}},
  \bibinfo{pages}{1921} (\bibinfo{year}{2005}).

\bibitem[{\citenamefont{Collins et~al.}(2002)\citenamefont{Collins, Gisin,
  Linden, Massar, and Popescu}}]{collins}
\bibinfo{author}{\bibfnamefont{D.}~\bibnamefont{Collins}},
  \bibinfo{author}{\bibfnamefont{N.}~\bibnamefont{Gisin}},
  \bibinfo{author}{\bibfnamefont{N.}~\bibnamefont{Linden}},
  \bibinfo{author}{\bibfnamefont{S.}~\bibnamefont{Massar}}, \bibnamefont{and}
  \bibinfo{author}{\bibfnamefont{S.}~\bibnamefont{Popescu}},
  \bibinfo{journal}{Phys. Rev. Lett.} \textbf{\bibinfo{volume}{88}},
  \bibinfo{pages}{040404} (\bibinfo{year}{2002}).

\bibitem[{\citenamefont{Finch}(2003)}]{finch}
\bibinfo{author}{\bibfnamefont{S.~R.} \bibnamefont{Finch}},
  \emph{\bibinfo{title}{Mathematical Constants}}
  (\bibinfo{publisher}{Cambridge}, \bibinfo{address}{New York},
  \bibinfo{year}{2003}).

\bibitem[{\citenamefont{Temperley and Fisher}(1961)}]{temperley}
\bibinfo{author}{\bibfnamefont{H.~N.~V.} \bibnamefont{Temperley}}
  \bibnamefont{and} \bibinfo{author}{\bibfnamefont{M.~E.}
  \bibnamefont{Fisher}}, \bibinfo{journal}{Philos. J.}
  \textbf{\bibinfo{volume}{6}}, \bibinfo{pages}{1061} (\bibinfo{year}{1961}).

\bibitem[{\citenamefont{Gagunashvili and Priezzhev}(1979)}]{gagun}
\bibinfo{author}{\bibfnamefont{N.~D.} \bibnamefont{Gagunashvili}}
  \bibnamefont{and} \bibinfo{author}{\bibfnamefont{V.~B.}
  \bibnamefont{Priezzhev}}, \bibinfo{journal}{Theor. Math. Phys.}
  \textbf{\bibinfo{volume}{39}}, \bibinfo{pages}{347} (\bibinfo{year}{1979}).

\bibitem[{\citenamefont{Akhtarshenas}()}]{iran2}
\bibinfo{author}{\bibfnamefont{S.~J.} \bibnamefont{Akhtarshenas}},
  \eprint{arXiv:0705.1965}.

\bibitem[{\citenamefont{Szarek et~al.}(2006)\citenamefont{Szarek, Bengtsson,
  and {\.Z}yczkowski}}]{sbz}
\bibinfo{author}{\bibfnamefont{S.}~\bibnamefont{Szarek}},
  \bibinfo{author}{\bibfnamefont{I.}~\bibnamefont{Bengtsson}},
  \bibnamefont{and}
  \bibinfo{author}{\bibfnamefont{K.}~\bibnamefont{{\.Z}yczkowski}},
  \bibinfo{journal}{J. Phys. A} \textbf{\bibinfo{volume}{39}},
  \bibinfo{pages}{L119} (\bibinfo{year}{2006}).

\bibitem[{\citenamefont{Gross et~al.}(1982)\citenamefont{Gross, Perry, and
  Yaffe}}]{gross}
\bibinfo{author}{\bibfnamefont{D.~J.} \bibnamefont{Gross}},
  \bibinfo{author}{\bibfnamefont{M.~J.} \bibnamefont{Perry}}, \bibnamefont{and}
  \bibinfo{author}{\bibfnamefont{L.~G.} \bibnamefont{Yaffe}},
  \bibinfo{journal}{Phys. Rev. D} \textbf{\bibinfo{volume}{25}},
  \bibinfo{pages}{330} (\bibinfo{year}{1982}).

\bibitem[{\citenamefont{Macdonald}(1995)}]{ig}
\bibinfo{author}{\bibfnamefont{I.~G.} \bibnamefont{Macdonald}},
  \emph{\bibinfo{title}{Symmetric Functions and Hall Polynomial}}
  (\bibinfo{publisher}{Oxford}, \bibinfo{address}{New York},
  \bibinfo{year}{1995}).

\bibitem[{\citenamefont{Pittenger and Rubin}(2002)}]{pittenger}
\bibinfo{author}{\bibfnamefont{A.~O.} \bibnamefont{Pittenger}}
  \bibnamefont{and} \bibinfo{author}{\bibfnamefont{M.~H.} \bibnamefont{Rubin}},
  \bibinfo{journal}{Lin. Alg. Applics.} \textbf{\bibinfo{volume}{346}},
  \bibinfo{pages}{47} (\bibinfo{year}{2002}).

\bibitem[{\citenamefont{Petz and Sud{\'a}r}(1996)}]{petzsudar}
\bibinfo{author}{\bibfnamefont{D.}~\bibnamefont{Petz}} \bibnamefont{and}
  \bibinfo{author}{\bibfnamefont{C.}~\bibnamefont{Sud{\'a}r}},
  \bibinfo{journal}{J. Math. Phys.} \textbf{\bibinfo{volume}{37}},
  \bibinfo{pages}{2662} (\bibinfo{year}{1996}).

\bibitem[{\citenamefont{Batle et~al.}(2006)\citenamefont{Batle, Casas,
  Plastino, and Plastino}}]{batlecasas1}
\bibinfo{author}{\bibfnamefont{J.}~\bibnamefont{Batle}},
  \bibinfo{author}{\bibfnamefont{M.}~\bibnamefont{Casas}},
  \bibinfo{author}{\bibfnamefont{A.}~\bibnamefont{Plastino}}, \bibnamefont{and}
  \bibinfo{author}{\bibfnamefont{A.~R.} \bibnamefont{Plastino}},
  \bibinfo{journal}{Phys. Lett. A} \textbf{\bibinfo{volume}{353}},
  \bibinfo{pages}{161} (\bibinfo{year}{2006}).

\bibitem[{\citenamefont{Batle et~al.}(2002)\citenamefont{Batle, Plastino,
  Casas, and Plastino}}]{batleplastino1}
\bibinfo{author}{\bibfnamefont{J.}~\bibnamefont{Batle}},
  \bibinfo{author}{\bibfnamefont{A.~R.} \bibnamefont{Plastino}},
  \bibinfo{author}{\bibfnamefont{M.}~\bibnamefont{Casas}}, \bibnamefont{and}
  \bibinfo{author}{\bibfnamefont{A.}~\bibnamefont{Plastino}},
  \bibinfo{journal}{Phys. Lett. A} \textbf{\bibinfo{volume}{298}},
  \bibinfo{pages}{301} (\bibinfo{year}{2002}).

\bibitem[{\citenamefont{Verstraete et~al.}(2001)\citenamefont{Verstraete,
  Audenaert, and Moor}}]{ver}
\bibinfo{author}{\bibfnamefont{F.}~\bibnamefont{Verstraete}},
  \bibinfo{author}{\bibfnamefont{K.}~\bibnamefont{Audenaert}},
  \bibnamefont{and} \bibinfo{author}{\bibfnamefont{B.~D.} \bibnamefont{Moor}},
  \bibinfo{journal}{Phys. Rev. A} \textbf{\bibinfo{volume}{64}},
  \bibinfo{pages}{012316} (\bibinfo{year}{2001}).

\bibitem[{\citenamefont{Hildebrand}(2007)}]{roland}
\bibinfo{author}{\bibfnamefont{R.}~\bibnamefont{Hildebrand}},
  \bibinfo{journal}{Phys. Rev. A} \textbf{\bibinfo{volume}{76}},
  \bibinfo{pages}{052325} (\bibinfo{year}{2007}).

\end{thebibliography}

\end{document}